\documentclass{ws-mplb}

\begin{document}

\markboth{Peter B. Weichman}{Dirty Bosons: Twenty Years Later}

%
\catchline{}{}{}{}{}
%

\title{Dirty Bosons: Twenty Years Later}

\author{\footnotesize Peter B. Weichman}

\address{BAE Systems, Advanced
Information Technologies, \\
6 New England Executive Park, Burlington,
MA 01803, USA \\
peter.weichman@baesystems.com}

%

\maketitle

\begin{history}
\received{(Day Month Year)} \revised{(Day Month Year)}
\end{history}

\begin{abstract}

A concise, somewhat personal, review of the problem of superfluidity
and quantum criticality in regular and disordered interacting Bose
systems is given, concentrating on general features and important
symmetries that are exhibited in different parts of the phase
diagram, and that govern the different possible types of critical
behavior.  A number of exact results for various insulating phase
boundaries, which may be used to constrain the results of numerical
simulations, can be derived using large rare region type arguments.
The nature of the insulator-superfluid transition is explored
through general scaling arguments, exact model calculations in one
dimension, numerical results in two dimensions, and approximate
renormalization group results in higher dimensions. Experiments on
$^4$He adsorbed in porous Vycor glass, on thin film superconductors,
and magnetically trapped atomic vapors in a periodic optical
potential, are used to illustrate many of the concepts.

\end{abstract}


\section{Introduction}
\label{sec:intro}

It has been approximately 20 years now since the flowering of
interest in the dirty boson problem. The problem is defined,
generally, as the nature of the insulating and conducting phases,
and the phase transitions between them, in a system of interacting
bosons in a random potential at zero temperature.  A phase
transition occurring at zero temperature, as a function of some
auxiliary control parameter, such as density or magnetic field, is
known as a \emph{quantum phase transition} (QPT) since the particle
dynamics are provided purely by the fluctuations in the ground state
wavefunction.\footnote{At finite temperature, through standard
arguments (see, e.g., Ref.\ \refcite{Hertz1976}) quantum mechanics
is irrelevant near criticality, and the physics reduces to that of
the corresponding classical model---the XY-model in the case of
superfluid $^4$He. The effects of disorder are then governed by the
well known Harris criterion (Ref.\ \refcite{Harris1974}). For
discussion specific to bosons on a disordered substrate, and further
references, see Ref.\ \refcite{WF1986}.} In a Bose system the
conducting phase is believed always to be a superfluid (SF) (see,
e.g., Ref.\ \refcite{Leggett1973}), and this strongly distinguishes
the dirty boson problem from the corresponding Anderson localization
and metal-insulator transition problems in Fermi
systems.\cite{LR1985,BK1994} In particular, the conducting phase has
a natural order parameter, absent in an ordinary metal, associated
with the off-diagonal long range order of the superfluid. Moreover,
in the absence of Pauli exclusion, repulsive interactions are
essential in forestalling condensation of a macroscopic number of
particles into the single lowest localized free particle eigenstate
of the random potential. There is therefore no sensible
noninteracting limit about which to perturb to study the physical
finite density phases. Thus, although there are some analogies
between the Fermi and Bose phenomena, the underlying physics is very
different, and the two problems require entirely different
theoretical approaches to their solution.

\begin{figure}[th]

\centerline{\psfig{file=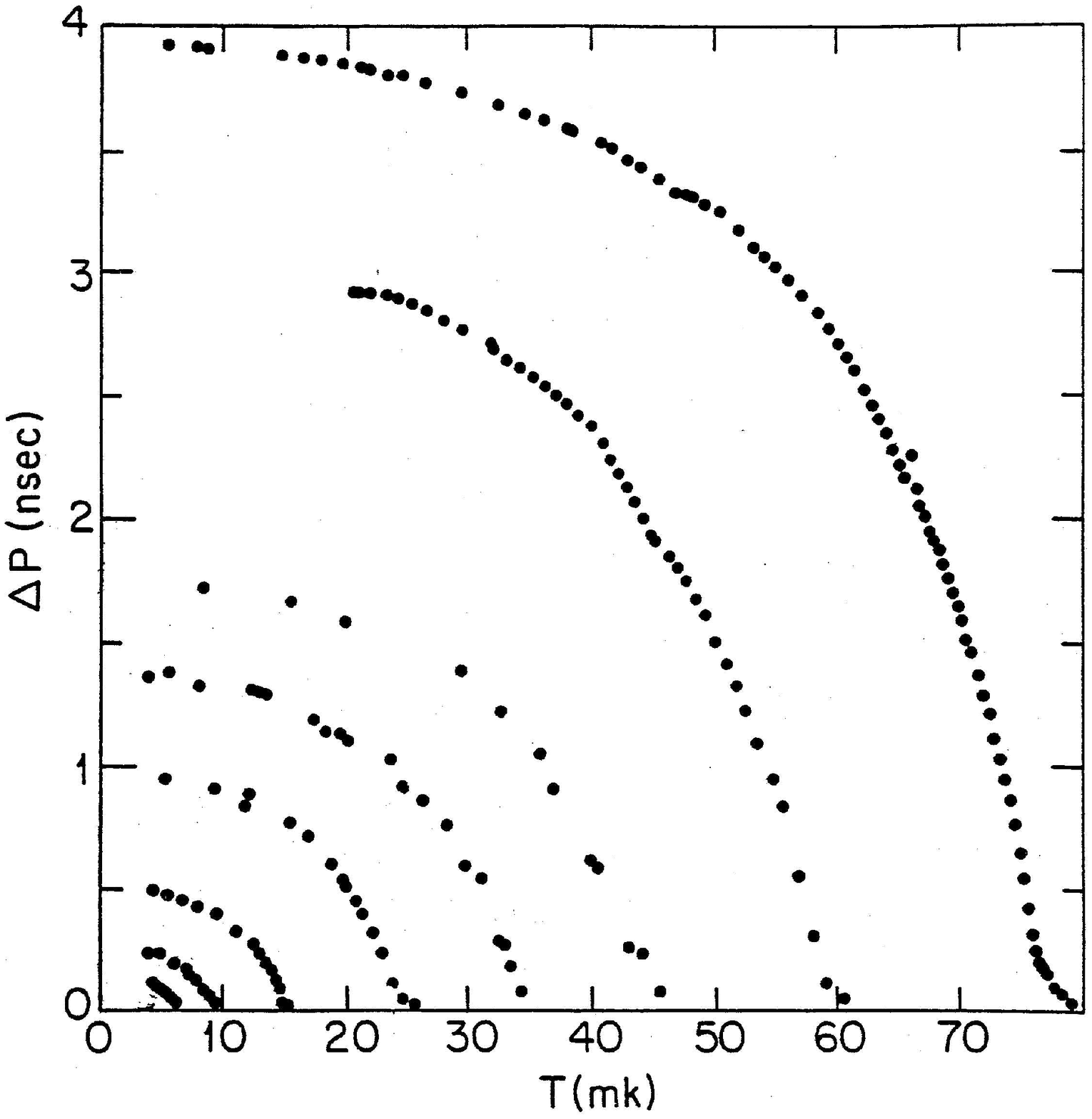,width=2.35in} \quad
\psfig{file=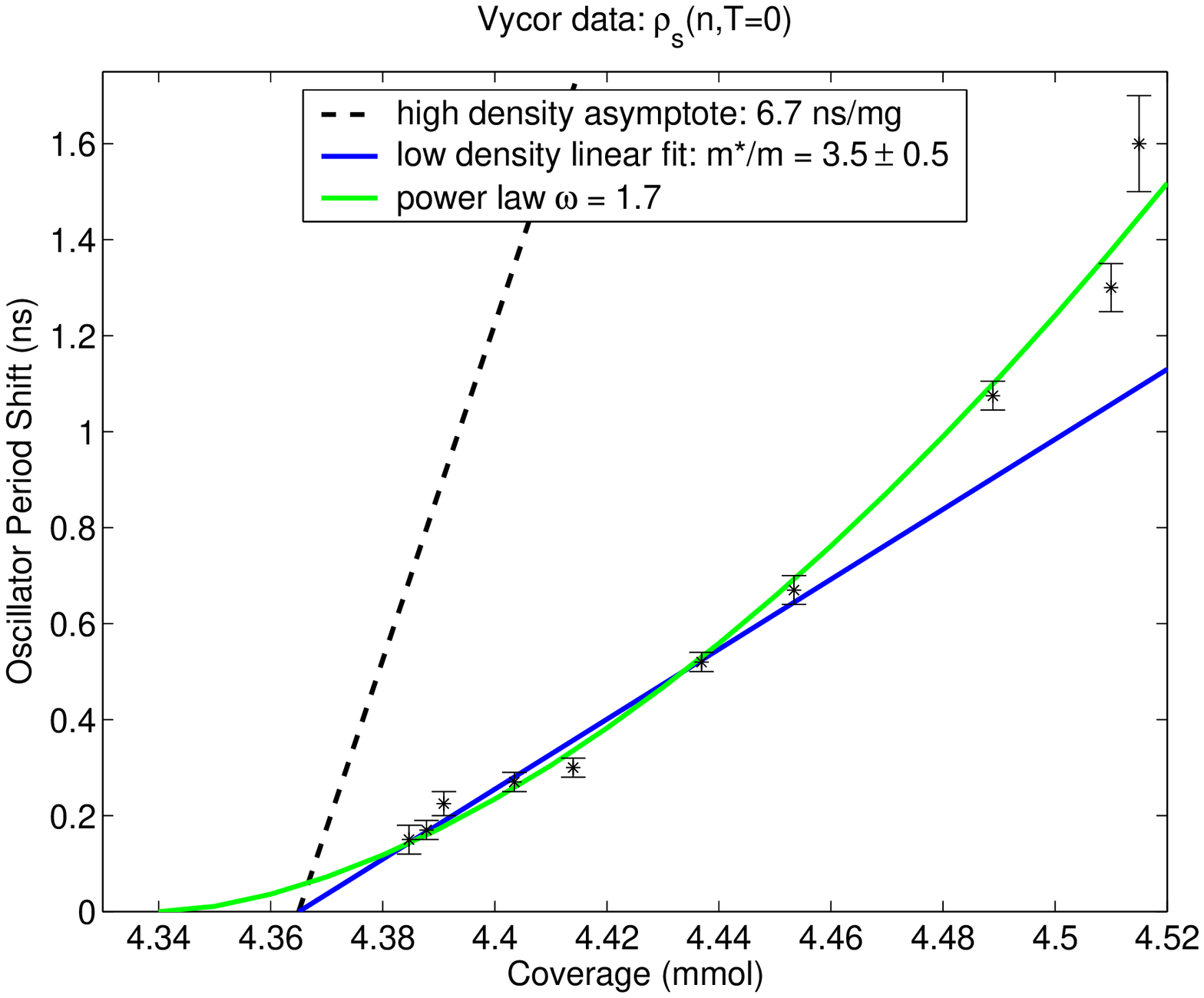,width=2.35in}}

\centerline{\psfig{file=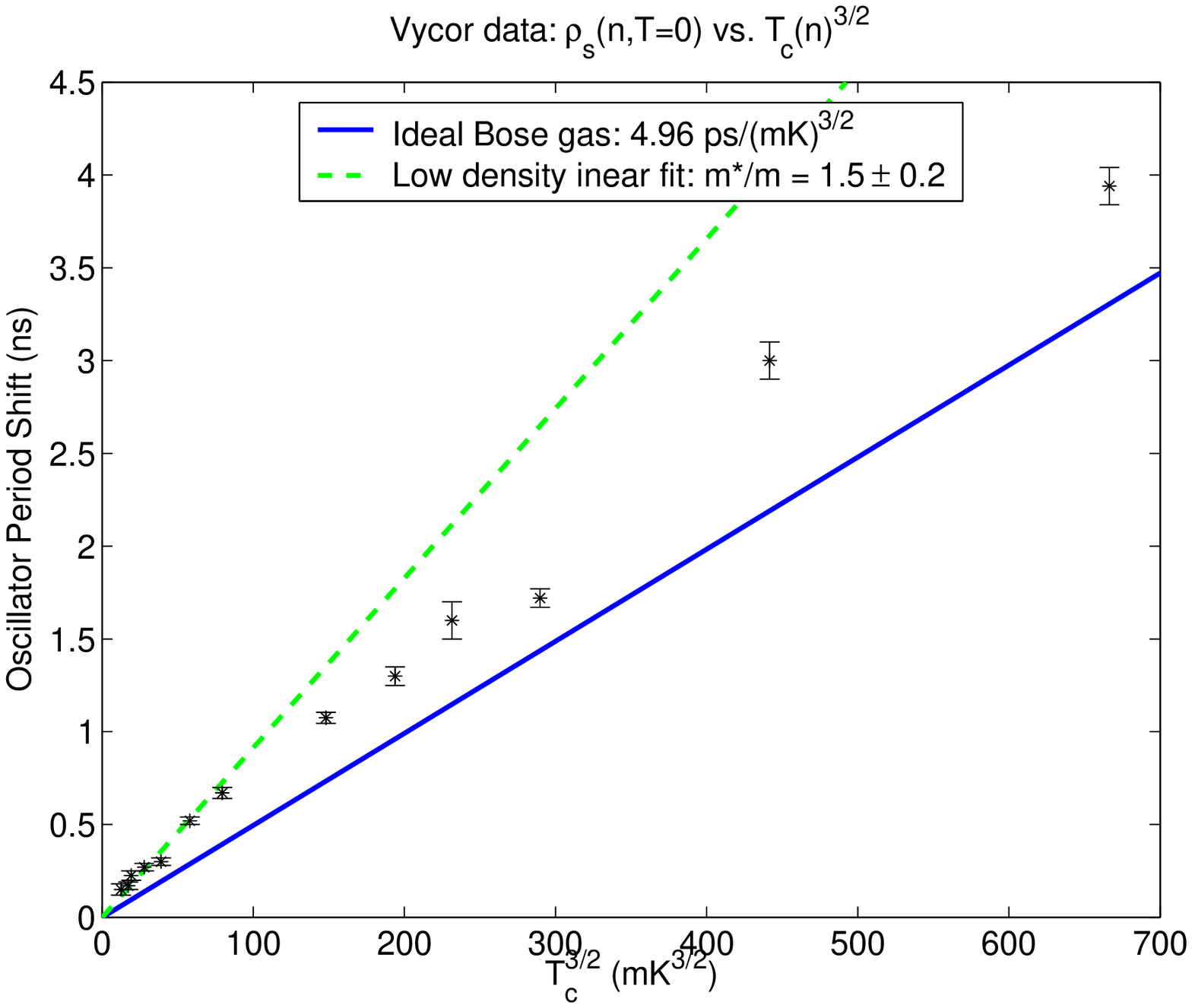,width=2.35in} \quad
\psfig{file=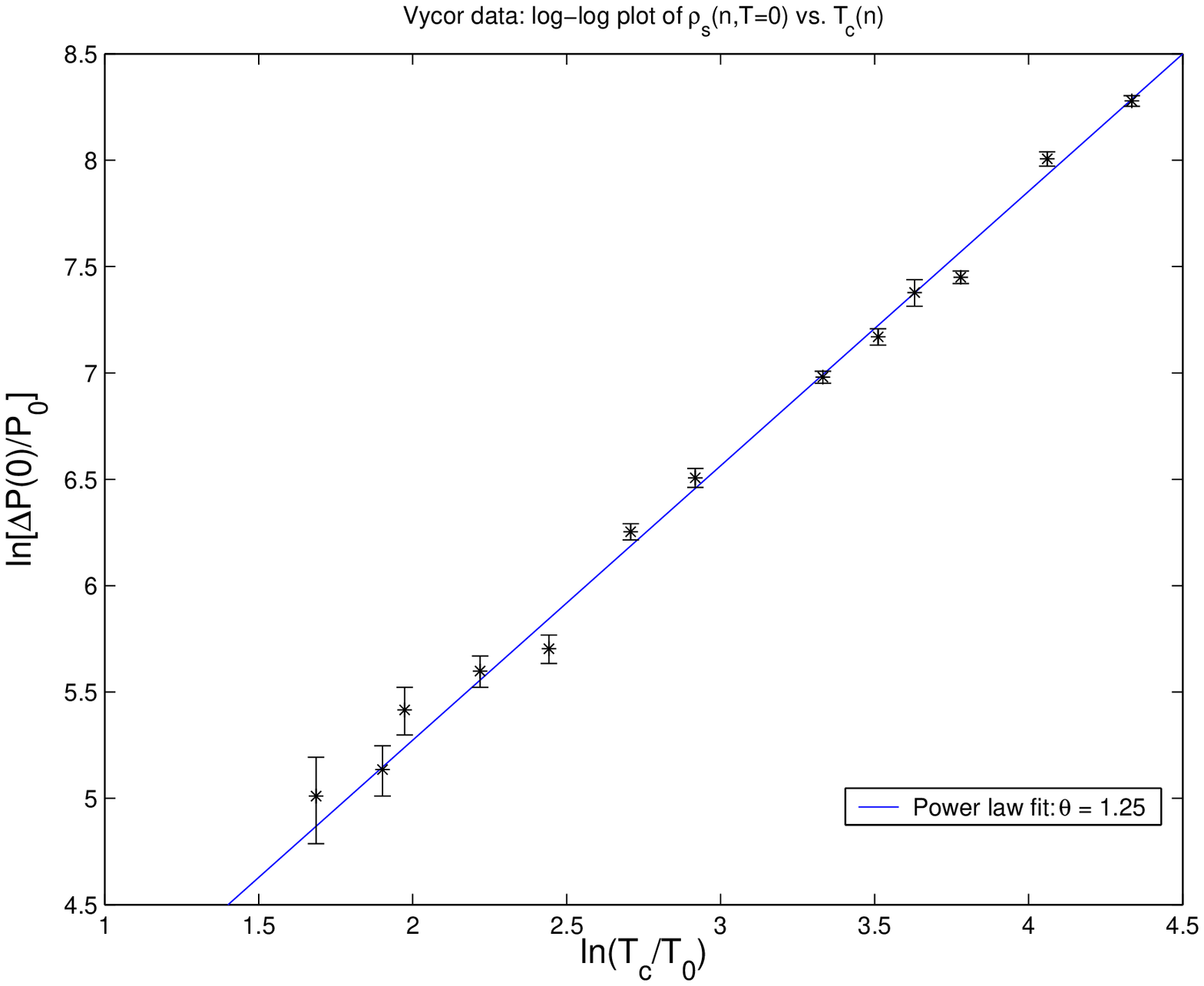,width=2.35in}}


\caption{Superfluidity of helium in Vycor. \textbf{Upper left:}
Superfluid density (proportional to the torsional oscillator period
shift $\Delta P$ due to the decoupling of the superfluid from the
oscillating substrate) data for $^4$He adsorbed in porous Vycor
glass. The different curves correspond to different fixed values of
the overall filling. The trend with decreasing coverage $\rho$,
towards a more linear, less steep, onset just below the transition
temperature $T_c(\rho)$, is evident to the eye. \textbf{Upper
right:} Extrapolated zero temperature superfluid density vs.\
coverage, with fits to ideal Bose gas and QPT-motivated functional
forms. \textbf{Lower left:} Extrapolated zero temperature superfluid
density vs.\ $T_c$ compared to ideal gas models. \textbf{Lower
right:} QPT-motivated power-law fit to the zero temperature
superfluid density vs.\ $T_c$ data.}

\label{fig:hevycor}

\end{figure}

\subsection{Some ``ancient'' history}
\label{sec:history}

Perhaps the earliest relevant experiments, which certainly were key
to motivating early work on the dirty boson problem,\cite{FWGF} were
on superfluidity of very thin films of $^4$He adsorbed in porous
Vycor glass.\cite{Reppy1983} Vycor glass has a highly connected,
40\% open (well above the percolation threshold), 3D sponge-like
structure with pore sizes in the 4--8 nm range. In an interesting
historical reversal, the experiments themselves were motivated not
by disorder effects, but by the claimed earliest experimental
evidence for an essentially ideal Bose gas, a dozen years in advance
of its observation in magnetically trapped atomic
vapors!\cite{BEC1995a,BEC1995b}  Of course, the Vycor system is not
nearly as ``clean'' as the atomic system (in senses that should
become clear below), so the claim itself is not as clean as might be
desired.\footnote{The essential role of Vycor is that it acts to
screen the long-range attractive part of the helium-helium pair
potential, which in the bulk causes the vapor to condense into a
dense fluid, preempting observation of the dilute regime in pure,
bulk $^4$He.}

Some of the Vycor superfluid density ($\rho_s$) vs.\ temperature
data\cite{Reppy1983} are reproduced in the upper left panel of Fig.\
\ref{fig:hevycor}. The ideal Bose glass claim emerged from an
examination of the evolution of the shapes of the $\rho_s(T;\rho)$
profiles in the vicinity of the transition $T_c(\rho)$ as the
overall helium density $\rho$ was reduced. It was
observed\cite{Reppy1983} that the exponent $\upsilon$ describing the
superfluid onset power law, $\rho_s \sim |t|^\upsilon$ below $T_c$,
where $t = (T-T_c)/T_c$ is the reduced temperature, showed a clear
crossover from the classic bulk value, $\upsilon \simeq 2/3$,
observed at higher coverages, toward the ideal Bose gas value,
$\upsilon_0 = 1$, at the lowest coverages. Quantitative theoretical
predictions for this crossover, based entirely on a clean,
non-disordered ``effective medium''
model,\cite{RSFW1984,WRFS1986,PBWthesis} were generally consistent
with this scenario.

Although the trends in the vicinity of $T_c$ support an ideal gas
interpretation, other elements of the data definitely do not, and
this will serve as motivation for the remainder of this article.
Motivated by QPT ideas, the remaining panels of Fig.\
\ref{fig:hevycor} show plots involving various zero temperature
quantities.\cite{PBWthesis}  In the upper right panel is plotted the
extrapolated zero temperature superfluid density $\rho_s(0;\rho)$
vs.\ $\rho$ (the horizontal axis should be divided by the 0.89
cm$^3$ total glass-plus-pore volume if one is fussy about units).
Zero temperature superfluidity evidently disappears for $\rho <
\rho_c \simeq 4.35$ mmol, corresponding to a coverage of about 1.5
monolayers (over the estimated 156 m$^2$ pore surface area). If one
interprets $\rho_c$ to be an inert background upon which the
superfluid component ``floats,'' then an ideal gas model would
predict a linear relation $\rho_s(0;\rho) \propto \rho-\rho_c$.
Evidently, one may force a linear fit to the lowest density data by
adjusting the slope using an effective mass free parameter. The
fitted value $m^*/m \simeq 3.5$ is at least consistent with one's
intuition that interaction with the substrate would tend to increase
the mass. However, a more convincing fit, over a much wider range,
is obtained with a superlinear fit $\rho_s(0;\rho) \propto
(\rho-\rho_c)^\omega$, which serves to define our first QPT critical
exponent $\omega \simeq 1.7 \pm 0.3$.

The bottom two panels of Fig.\ \ref{fig:hevycor} compare the rates
at which $\rho_s(0;\rho)$ and $T_c(\rho)$ vanish with $\rho$. An
ideal gas model predicts $T_c \propto \rho^{d/2}$, with spatial
dimensionality $d=3$ here. The lower left panel shows that one may
force such a proportionality over the lower coverage data, again
using an effective mass free parameter. The fact that the inferred
value $m^*/m \simeq 1.5$ is \emph{different} from the previous one
should lead to some discomfort. Once again, as seen in the log-log
plot of the same data in the right panel, a much more satisfactory
power-law fit $\rho_s(0;\rho) \propto T_c(\rho)^\theta$ is obtained,
with a second QPT critical exponent $\theta = 1.25 \pm
0.2$.\footnote{Why, over a large range of coverages, is the behavior
near $T_c$ consistent with an ideal Bose gas crossover, while over
the same range the zero temperature quantities show strong
deviations? The answer is not entirely clear but, firstly, critical
scaling near $T_c$ is entirely separate from QPT scaling near
$T=0$---for a detailed discussion of this in the context of the
clean system, see Ref.\ \refcite{W1989}.  Thus, disorder is expected
to have distinctly different impacts on the two ends of the
superfluid density profile. Second, as discussed in Ref.\
\refcite{WF1986}, its manufacturing process makes Vycor very uniform
on multi-pore scales, which aids the uniform effective medium
approximation which forms the basis of the ideal gas model. However,
one eventually expects to see strong deviations, as the disorder
begins to impact the ideal Bose gas physics, presumably at yet
smaller values of $\rho-\rho_c$ that are beyond experimental
resolution.}

\subsection{Outline}
\label{sec:outline}

With the helium in Vycor system serving as an introduction, the
outline of the remainder of this paper is as follows.  This review
is not intended to be comprehensive, and the author apologizes in
advance for his personal choices as to which developments deserve
emphasis. In Sec.\ \ref{sec:models} the theoretical approach to the
problem is introduced through the Bose-Hubbard and Josephson
junction lattice models. In Sec.\ \ref{sec:phases}, the phases,
phase diagrams, and accompanying phase transitions are motivated and
summarized. In Sec.\ \ref{sec:droplet}, exact results for the
incompressible phase boundaries, based on the existence of
exponentially rare, but arbitrarily large, disorder free regions,
are derived. In Sec.\ \ref{sec:pathint} the quantum model is mapped
onto a classical field theory Lagrangian. The various parameters
appearing in the latter allow one to transparently identify the
different possible symmetries of the model, and accompanying QPT
universality classes. In Sec.\ \ref{sec:scaling}, QPT critical
scaling phenomenology is introduced, which, in particular, allows
one to express the exponents $\omega$ and $\theta$ in terms of more
familiar ones. Scaling also predicts the existence of universal
sheet conductances in $d=2$, which has some experimental support.
Renormalization group and epsilon expansion approaches to detailed
model calculations are discussed in Sec.\ \ref{sec:rg}. The latter
are poorly controlled compared to their classical counterparts, but
nevertheless provide an compelling global picture of the stability,
bifurcation, and merging of critical fixed points as a function of
dimension that is consistent with the phenomenological scaling
arguments. The review is summarized and concluded in Sec.\
\ref{sec:conclude}.

\section{Bose-Hubbard and Josephson junction Models}
\label{sec:models}

In analogy to the Hubbard model of electronic propagation in a
crystalline solid, a useful starting point for the study of bosons
in a disordered medium is the Bose-Hubbard model of lattice bosons.
The Hamiltonian is
\begin{equation}
{\cal H}_B = -J \sum_{\langle i,j \rangle}
\hat a_i^\dagger (\hat a_j - \hat a_i)
- \mu \sum_i \hat n_i
+ \frac{1}{2} V \sum_i \hat n_i (\hat n_i - 1)
+ \sum_i u_i \hat n_i,
\label{2.1}
\end{equation}
in which $\hat a_i^\dagger, \hat a_i$ are boson creation and
annihilation operators on site $i$ (taken to lie on a regular
lattice, and obeying the usual Bose commutation relations $[\hat
a_i,\hat a_j^\dagger] = \delta_{ij}$), $\hat n_i = \hat a_i^\dagger
\hat a_i$ is the (non-negative integer) site number operator, the
first sum is over nearest neighbors, $J = \hbar^2/2m^* d^2$ is the
hopping amplitude, where $d$ is a measure of the pore diameter and
$m^*$ is a boson effective mass, $\mu$ is the chemical potential, $V
> 0$ the on-site repulsion, and $u_i$ the random site energies. In
this model each site represents a pore, and the microscopic
structure of the individual pores has been subsumed into the model
effective parameters. The same picture obtains for bosons hopping
between minima of the optical lattice potentials that are used to
perturb magnetically trapped atomic vapors.\cite{BEClatta,BEClattb}
Clearly, a more realistic model would include disorder in $J$, $V$,
and the lattice site positions, but, as in the fermion case, these
add nothing new to the basic physics implied by the random $u_i$.
One could also add higher order Hubbard bands, corresponding to
higher order intra-pore single particle excited states, but at low
temperatures, lacking Pauli exclusion, these will not be populated,
and can be ignored. The precise probability distribution of the
$u_i$ is not important, but it is useful to consider them as
independent from site to site, with an even distribution (in
particular, with mean zero, which normalizes the chemical
potential), and with support on a finite interval $-\Delta \leq u_i
\leq \Delta$, with the bound $\Delta$ serving as a free parameter.

A closely related model is the Josephson junction, or quantum rotor
model,
\begin{equation}
{\cal H}_J = -\sum_{\langle i,j \rangle}
J_{ij} \cos(\hat \phi_i - \hat \phi_j)
- \mu \sum_i \hat n_i
+ \frac{1}{2} V \sum_i \hat n_i^2
+ \sum_i u_i \hat n_i,
\label{2.2}
\end{equation}
in which $\hat n_i, \hat \phi_i$ are conjugate number and phase
operators, defined by the commutation relations $[\hat \phi_i,\hat
n_j] = i \delta_{ij}$. In this model $\hat n_i$ can take negative as
well as positive integer values. The mapping between the two models
is provided by identifying $\hat a_i^\dagger \leftrightarrow \hat
n_i^{1/2} e^{i\hat \phi_i}, \hat a_i \leftrightarrow  e^{-i\hat
\phi_i} \hat n_i^{1/2}$. This Hamiltonian originated as a model of
granular superconductors, with $\hat n_i$ representing the number of
Cooper pairs on grain $i$, and with Josephson couplings $J_{ij}$
between the phases of the neighboring grains. The hopping term
clearly lowers the energy when neighboring phases are aligned, and
the superfluid, or superconducting, state corresponds to the
appearance of long range order, and corresponding nonzero local
averages $\langle e^{i\hat \phi_i} \rangle \neq 0$ [corresponding to
the usual Bose condensate $\langle \hat a_i \rangle \neq 0$ in
(\ref{2.1})]. Quantitative agreement between the two models is
obtained in the limit of large mean site occupancy $n_i = \langle
\hat n_i \rangle \gg 1$, where number fluctuations can be neglected
in the hopping term and one identifies $J_{ij} \approx J \sqrt{n_i
n_j}$.

By separating amplitude and phase in this way the underlying
symmetries of the model, which are critical to identifying the
possible QPT universality classes, are much more transparently
exhibited. First note that under an integer translation $\hat n_i
\to \hat n_i + n_0$, which preserves the commutation relations, the
Hamiltonian transforms in the form,
\begin{equation}
{\cal H}_J(\mu) \to {\cal H}_J(\mu - n_0 V)
+ \left(-\mu n_0 + \frac{1}{2} V n_0^2 \right) N_L,
\label{2.3}
\end{equation}
where $N_L$ is the number of lattice sites. It follows that the
phase diagram is periodic in $\mu$ with period $V$ (see Fig.\
\ref{fig:josephson_phases}; the additive constant serves simply to
increase the overall density by $n_0$). Second, the particle-hole
transformation corresponds to the reflections,
\begin{equation}
\hat n_i \to - \hat n_i,\ \hat \phi_i \to -\hat \phi_i,
\label{2.4}
\end{equation}
which also preserves the commutation relations. Under this
transformation, ${\cal H}(\mu,\{u_i\}) \to {\cal H}(-\mu,\{-u_i\})$.
Together with the translation symmetry, this implies that at the
special values $\mu = n_0 V/2$, in the absence of site disorder,
$u_i \equiv 0$ (but arbitrary $J_{ij}$), the Hamiltonian is
invariant (up to an additive constant), implying the existence of a
special symmetry between particle and hole excitations. Furthermore,
for symmetrically distributed $u_i$, the transformed Hamiltonian is
\emph{statistically} identical to the original. In either case, the
phase diagram has an additional reflection symmetry about the points
$\mu = n_0 V/2$.

These special \emph{particle-hole symmetric} and \emph{statistically
particle-hole symmetric} points (which ${\cal H}_B$ can only exhibit
indirectly---see below) will play an important role in understanding
the critical behavior. Furthermore, the fact that the hopping is
unaffected by the transformation motivates consideration of disorder
in both $J_{ij} = J(1 + \delta J_{ij})$ (which serves to define an
overall control parameter $J$) and $u_i$ in this model.
Understanding the effects of breaking these symmetries (which
ultimately leads back to ${\cal H}_B$, but now understood in a
broader context) will turn out to be key to understanding the global
renormalization group flows (Sec.\ \ref{sec:rg}).

\begin{figure}[th]

\centerline{\raisebox{0.75in}
{\psfig{file=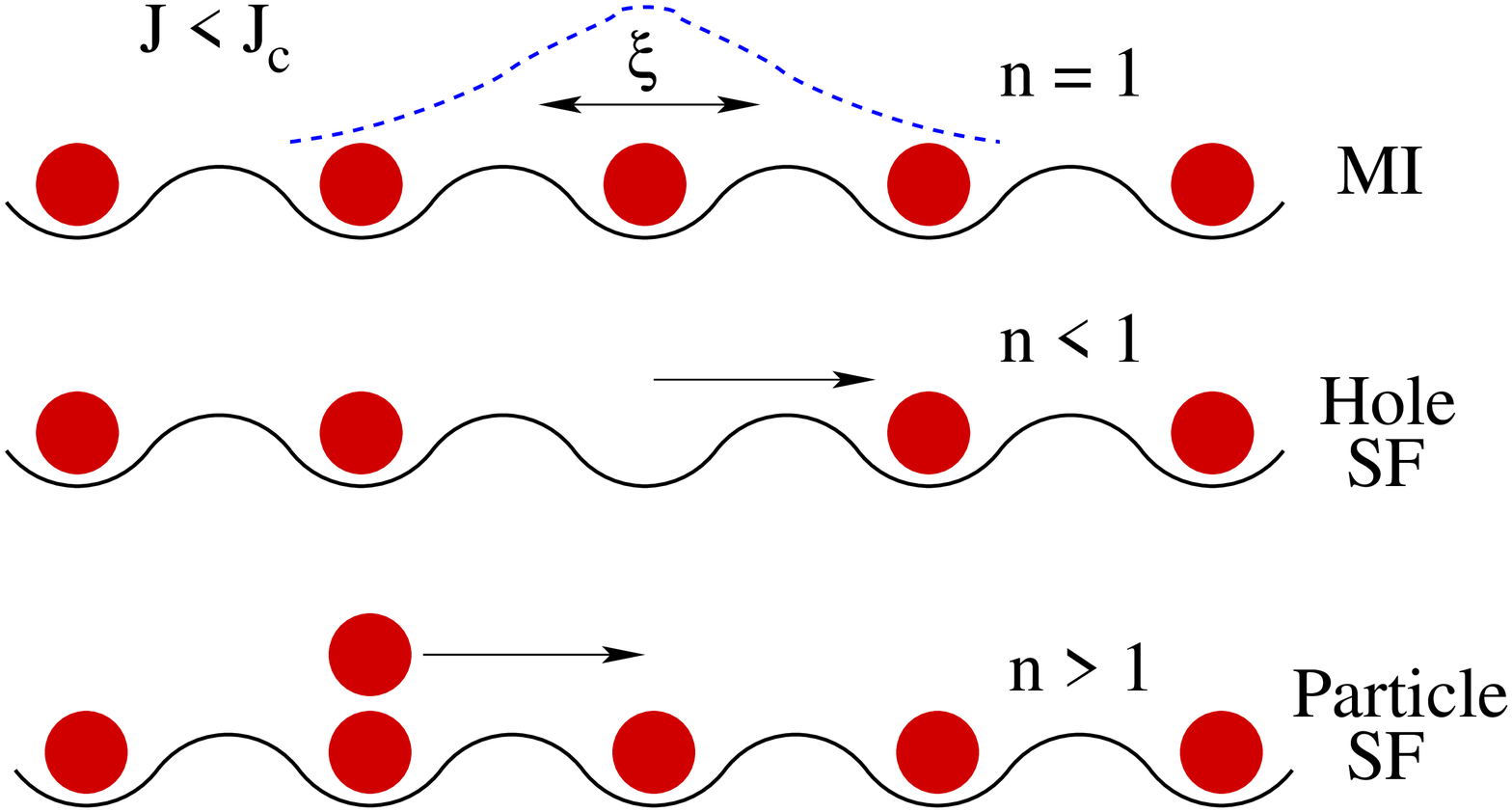,width=2.5in}}
\quad \psfig{file=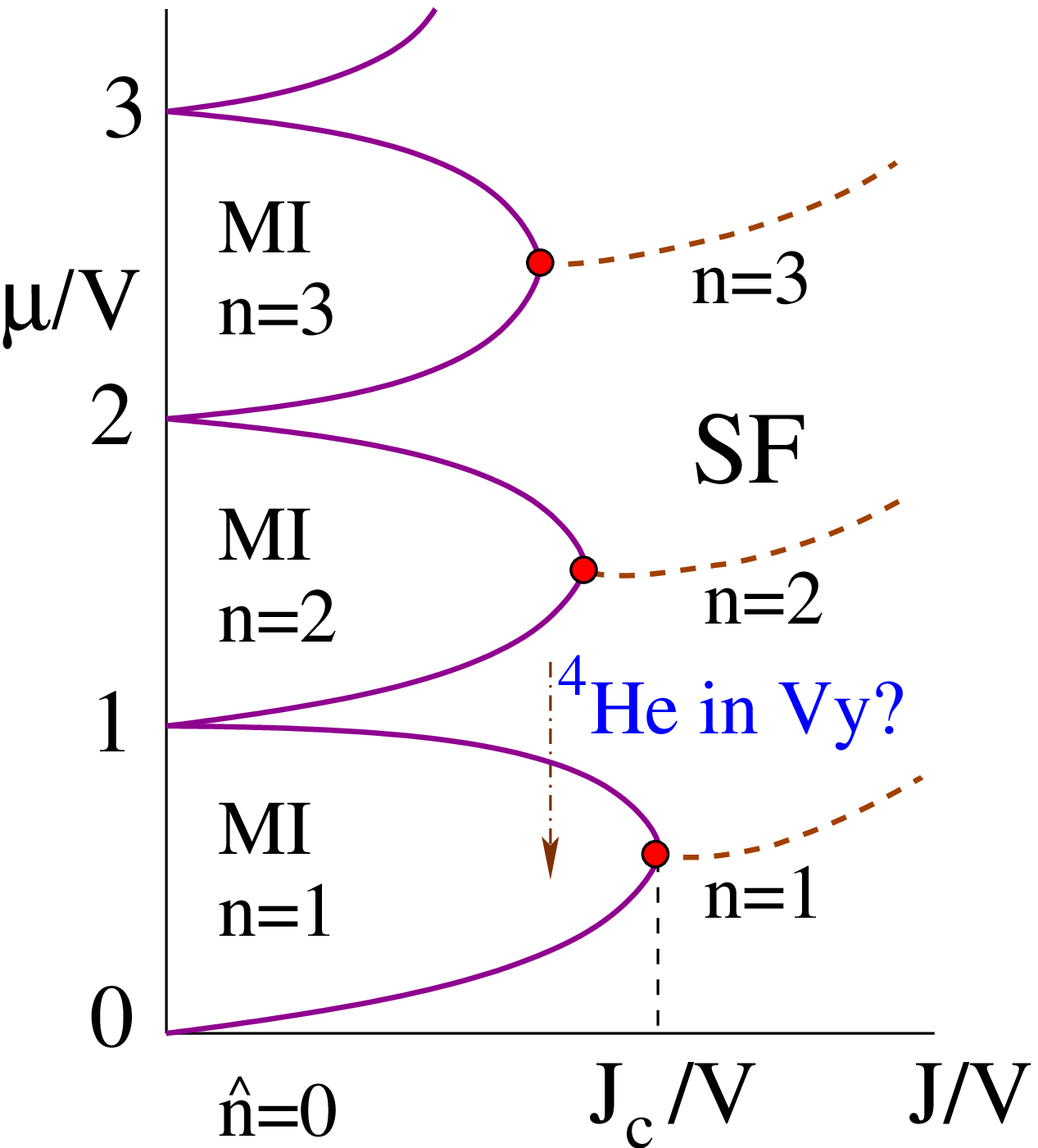,width=2.2in}}


\caption{\textbf{Left:} Schematic illustration of lattice bosons
near unit filling $n=1$ at some value of the hopping amplitude in
the interval $0 < J < J_c$. In the Mott insulating phase ($n=1$),
the effective wavefunction of each particle spreads only a finite
distance $\xi(J)$, and the state is insulating.  For $n > 1$ ($n <
1$), the extra particles (holes) travel freely within the
essentially inert background, and the state is superfluid for
arbitrarily small $|n-1|$. \textbf{Right:} Associated phase diagram
with Mott insulating (MI) and superfluid (SF) phases. The MI are
incompressible, with fixed integer filling over a finite range of
$\mu$. The SF is compressible, with integer filling only along some
set of lines extending from the Mott lobe tips. The phase transition
through the tip, where a ``hidden'' particle-hole symmetry is
asymptotically restored, is in the universality class of the
$(d+1)$-dimensional classical XY model. Elsewhere, the transition is
equivalent to onset of superfluidity at zero density in a continuum
dilute Bose gas.  In this sense, as indicated, ``periodic Vycor''
would indeed produce the desired ideal Bose gas crossover upon
approach to integer filling.}

\label{fig:hopping}
\end{figure}

\section{Phases and phase diagrams}
\label{sec:phases}

By considering the energy balance between the various terms in
(\ref{2.1}) and (\ref{2.2}) one can understand the basic topology of
the phase diagrams. First the results for the Bose-Hubbard model
(\ref{2.1}) will be motivated and presented. Additional features
special to the Josephson junction model (\ref{2.2}) will then be
discussed. Detailed calculations can be found in the literature: the
original mean field predictions for the phase diagram were made in
Ref.\ \refcite{FWGF}; subsequent perturbation calculations can be
found in Ref.\ \refcite{FM1994,NFM1999}, and numerical simulations
can be found, for example, in Refs.\
\refcite{SBZ1991,KTC1991,WSGY1994,ZKCG1995,AS2003,PS2004}.

\subsection{Clean system}

We begin with the clean, non-disordered system, $u_i \equiv 0$,
corresponding to bosons moving in a perfectly periodic background
potential. The left panel of Fig.\ \ref{fig:hopping} sketches the
essential physics. If there is no hopping, $J = 0$, then the sites
decouple, the site occupancies are good quantum numbers, and there
are exactly $n_0$ particles on each site for $n_0 < \mu/V < n_0+1$.
There is a unit jump in filling at each integer value of $\mu/V$.
For finite $J$, particles are no longer confined to a single site,
but may wander to neighboring sites. However, for small $J/V$ the
mutual repulsion produces an energy gap of order $V$ whenever two
particles occupy the same site. So, although virtual exchanges do
occur, the net effective distance $\xi(J,n_0)$ a given particle
wanders remains finite. There is a corresponding energy gap
$\mu_\pm(J,n_0) = \pm[E(J,n_0 N_L \pm 1)-E(J,n_0 N_L)]$, where
$E(J,N)$ is the ground state energy for $N$ particles, for adding
($+$) or removing ($-$) a particle. Thus, for chemical potential in
the interval $\mu_-(J,n_0) < \mu < \mu_+(J,n_0)$, the ground state
wavefunction is insensitive to $\mu$, and, in particular, the
density remains fixed at $n_0$. This incompressible phase is known
as a Mott insulator (MI), analogous to the corresponding electron
half-filled band insulators, and the resulting lobe structure of the
phase diagram is sketched in right panel of Fig.\ \ref{fig:hopping}.
All of these notions can be confirmed through direct perturbation
theory in $J/V$.\cite{FM1994,NFM1999}

Only for sufficiently large $J$ is the hopping sufficiently vigorous
to overcome the repulsion:  at a critical value $J \to J_c(n_0)$ the
hopping range diverges, $\xi \to \infty$, and the gap closes,
$\varepsilon_\pm \to 0$. For $J > J_c$ one enters the superfluid
phase. One may define a quantum critical exponent correlation length
exponent $\nu$ via $\xi \sim |J-J_c|^{-\nu}$, and dynamical exponent
$z$ via $\varepsilon_\pm \sim |J-J_c|^{z\nu}$. It will be shown in
Sec.\ \ref{sec:pathint} that this transition, at which, as alluded
to above, an \emph{effective} particle-hole symmetry is restored, is
in the universality class of the classical $(d+1)$-dimensional XY
model: $\nu = \nu_{d+1}^\mathrm{XY}$, $z=1$. Thus, for example, in
$d=2$, $\nu_3^\mathrm{XY} \simeq 0.671$ should be the same as that
measured at the $^4$He lambda transition.

Between Mott lobes, $\mu_+(J,n_0) < \mu < \mu_-(J,n_0+1)$, the
density varies continuously between the two integer values, $n_0 <
n(\mu,J) < n_0+1$. Near the Mott lobe boundaries one may think of
the difference $n-n_0$ as extra particles (or $n_0+1-n$ as extra
holes) propagating atop the Mott phase background. Now lacking a
barrier to hopping between sites, these particles (or holes) form a
superfluid (lower left part of Fig.\ \ref{fig:hopping}). Although
the notion of coexisting insulating and superfluid components is
misleading (bosons are identical particles, and there is strong
exchange between the two groups), the effective theory near the
phase boundary (analogous, perhaps, to the Fermi liquid theory of
effectively free electron excitations near the Fermi surface) is
indeed that of a dilute Bose gas. The exchange with the background
insulator gives rise to a strongly renormalized (especially near
$J_c$) effective single particle mass $m^*$. In this sense, if Vycor
had perfectly periodic pores, experiments would indeed see a
crossover to dilute Bose gas behavior, with $\rho_c = n_0/d^3$
corresponding to the density at integer filling. Perhaps the closest
experimental realization is that of trapped atomic vapors in an
additional imposed optical lattice.\cite{BEClatta,BEClattb} However,
although both MI and SF phases have been detected in these
experiments, trap inhomogeneities, finite size effects, and other
measurement constraints, are such that detailed critical phenomena
are not yet resolvable.

As stated, the very different critical behaviors at the
``commensurate'' transition at $J_c$, and the line of
``incommensurate'' transitions below $J_c$ are indicative of a
``hidden'' particle-hole symmetry at the former. The lack of an
exact symmetry in ${\cal H}_B$ means that this symmetry is
effectively restored at the nontrivial point where $\mu_+(J_c,n_0) =
\mu_-(J_c,n_0)$ (analogous to the asymptotic restoration of the
underlying Ising up-down symmetry of density fluctuations near a
liquid-vapor critical point). During the transition to superfluidity
at integer filling, particle and hole excitations exist in equal
numbers, tunneling through each other, and achieve superfluidity
simultaneously. Below $J_c$ the transition takes place by feeding
extra particles or holes into the system, explicitly breaking this
symmetry, and only the predominant excitation achieves
superfluidity.

\begin{figure}[th]

\centerline{\raisebox{0.5in}
{\psfig{file=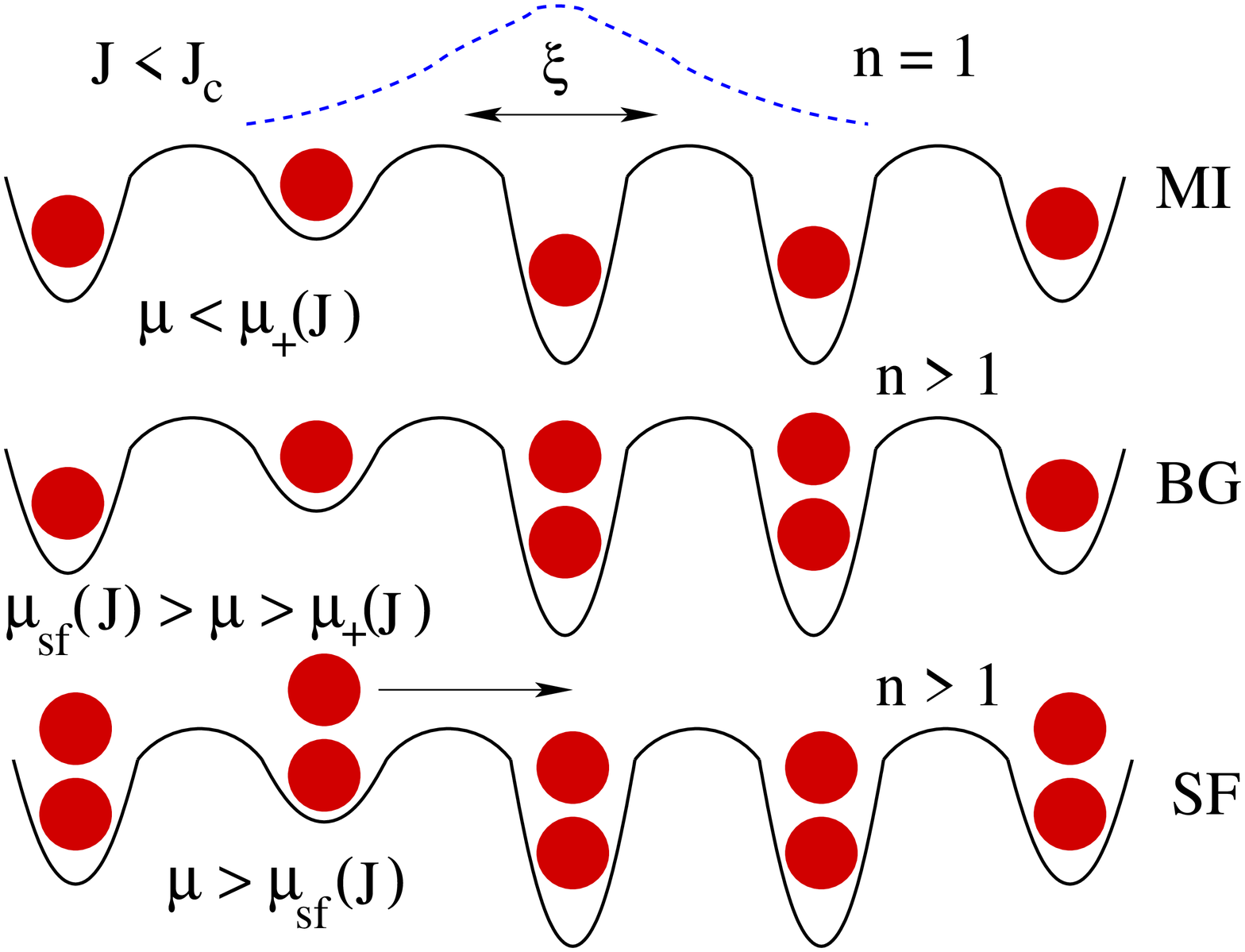,width=2.4in}}
\quad \psfig{file=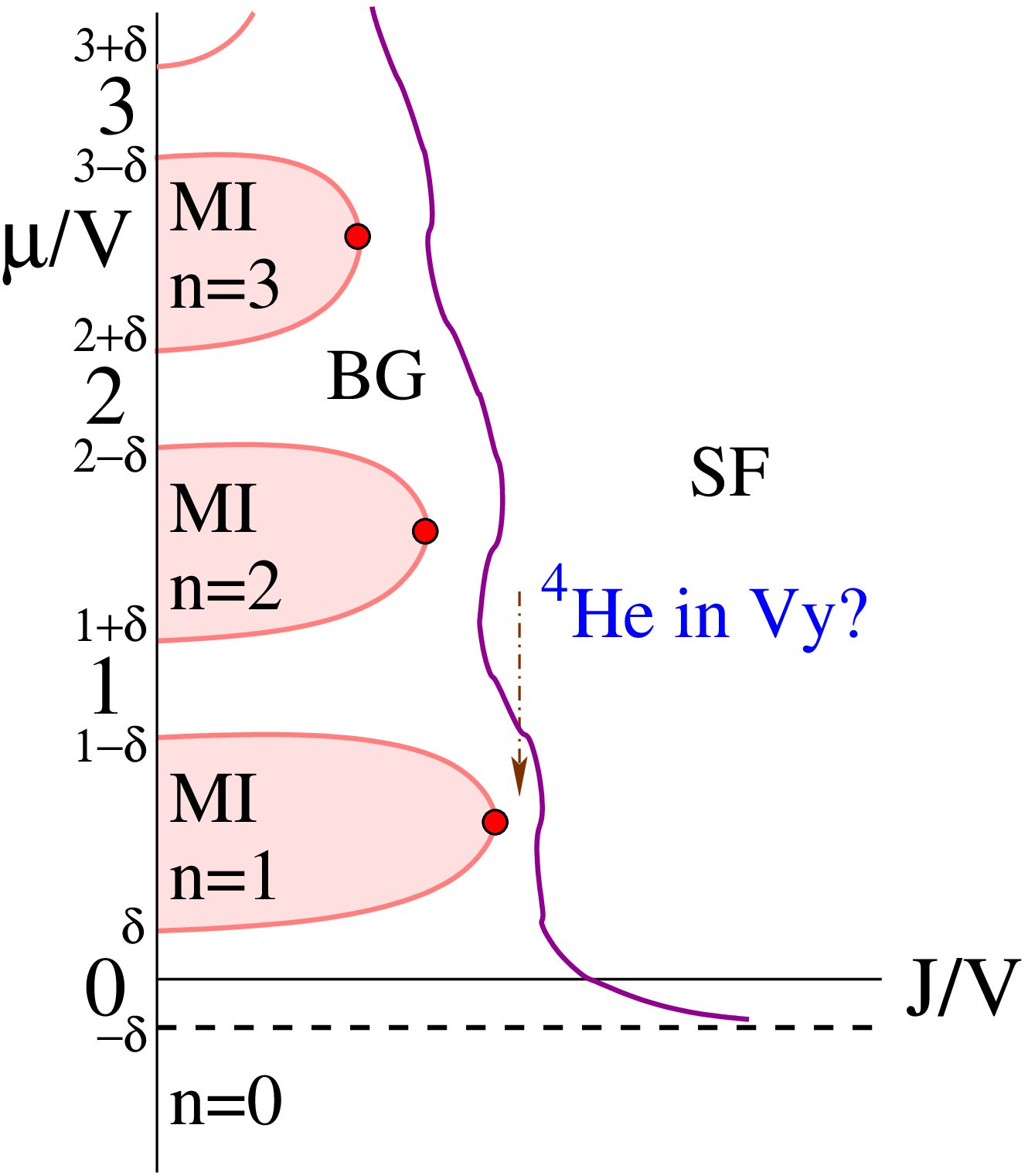,width=2.3in}}


\caption{\textbf{Left:} Schematic illustration of lattice bosons
near unit filling in the presence of bounded site disorder.  If the
disorder is not too strong ($\delta \equiv \Delta/V < \frac{1}{2}$),
there is still a (shrunken) Mott lobe with a finite energy gap for
adding or removing particles. Unlike in the pure case (Fig.\
\ref{fig:hopping}), superfluidity is not generated immediately
outside this lobe. For sufficiently small $n-1$, the additional
particles are Anderson localized by the residual random background
potential of the effectively inert layer. A finite compressibility
distinguishes this new insulating Bose glass (BG) phase from the MI.
The superfluid critical point $\mu_\mathrm{sf}(J)$ occurs only once
the added particles have sufficiently smoothed the background
potential that its effective lowest lying single particle states
become extended. \textbf{Right:} Associated phase diagram, with MI,
BG and SF phases. The transition to superfluidity is always from the
BG phase, is in the same universality class along the entire
transition line, and is ultimately the correct description of helium
in Vycor.}

\label{fig:ranhopping}
\end{figure}

\subsection{Disordered system}

Consider now the addition of site disorder.  The left panel of Fig.\
\ref{fig:ranhopping} motivates the existence of a third phase, the
Bose glass (BG) phase,\footnote{The term ``Bose glass'' may have
originated in the early work of Ref.\ \refcite{HFA1979}, who studied
a Hartree-Fock approximation to (\ref{2.1}). It transpires that such
an effective single-particle treatment of the interactions is a very
poor approximation for disordered bosons, and their main conclusions
were disputed in Ref.\ \refcite{BM1982}.} that intervenes between
the Mott and superfluid phases.\cite{FWGF} Beginning again with the
$J=0$ limit, for sufficiently bounded disorder, $\Delta < V/2$,
there remains an interval $(n_0 - 1)V + \Delta < \mu < n_0 V -
\Delta$ over which every site still has exactly $n_0$ particles.
However, for $\mu$ just outside this interval, some sites with have
an extra particle (or hole), and the fraction of such sites will
vary continuously with $\mu$---even at $J=0$ the state is now
compressible.

\begin{figure}[th]

\centerline{\psfig{file=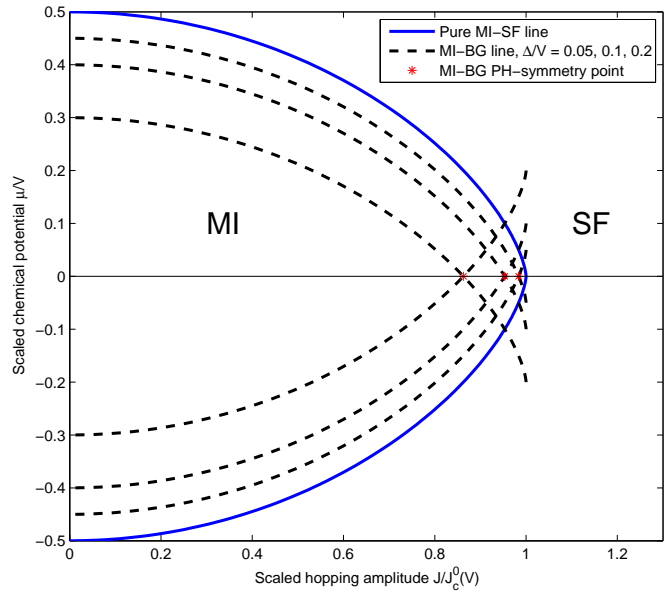,width=4.0in}}

\caption{Sketch of the Mott lobe boundaries as a function of the
disorder (adapted from Ref.\ \protect\refcite{WM08}). For
simplicity, we consider here the $n_0 = 0$ lobe of the Josephson
junction model, for which $\mu_+ = -\mu_-$. The phase boundaries are
simply those of the pure system translated inwards by the disorder
distribution half-width $\Delta$: $\mu_\pm(J,\Delta) = \mu_\pm(J,0)
\mp \Delta$. The result, for example, is a corner, or slope
discontinuity, at the Mott lobe tip $J_M(\Delta)$ satisfying
$\mu_+(J_M,0)-\mu_-(J_M,0) = 2\Delta$ (marked by stars). }

\label{fig:mibg_bdy}
\end{figure}

For small $J$, there will still be an interval
$\mu_c^-(J,\Delta,n_0) < \mu < \mu_c^+(J,\Delta,n_0)$ where mutual
repulsion continues to the dominate the disorder, and the
incompressible Mott lobe, though shrunken, still survives (right
panel of Fig.\ \ref{fig:ranhopping}). A rare region argument will be
used in Sec.\ \ref{sec:droplet} to show that $\mu_\pm(J,\Delta,n_0)
= \mu_\pm(J,0,n_0) \mp \Delta$ are simple translates of those of the
clean system (Fig.\ \ref{fig:mibg_bdy}).

The compressible phase just outside the Mott lobe, however, can no
longer be a superfluid. The extra particles (or holes) propagating
atop the background Mott phase now encounter a background random
potential, and the usual Anderson arguments\cite{LR1985} show that
the low energy effective single particle states must be localized,
and the state therefore remains insulating. Only after a sufficient
density of particles has been added, $\mu > \mu_\mathrm{sf}(J)$, is
the residual random potential sufficiently smooth that that the
lowest lying state becomes extended, and the transition to
superfluidity takes place. Note the key role here of the pair
interactions that allow particles to gradually ``fill in'' the
deeper minima in the background potential.

Note that for $\Delta > V/2$ (which includes unbounded, e.g.,
Gaussian, disorder) the Mott lobes are entirely destroyed, and only
the Bose glass and superfluid phases survive. This is presumably the
case for helium in Vycor, where the pore irregularities are too
strong to permit a Mott phase.

\begin{figure}[th]

\centerline{\psfig{file=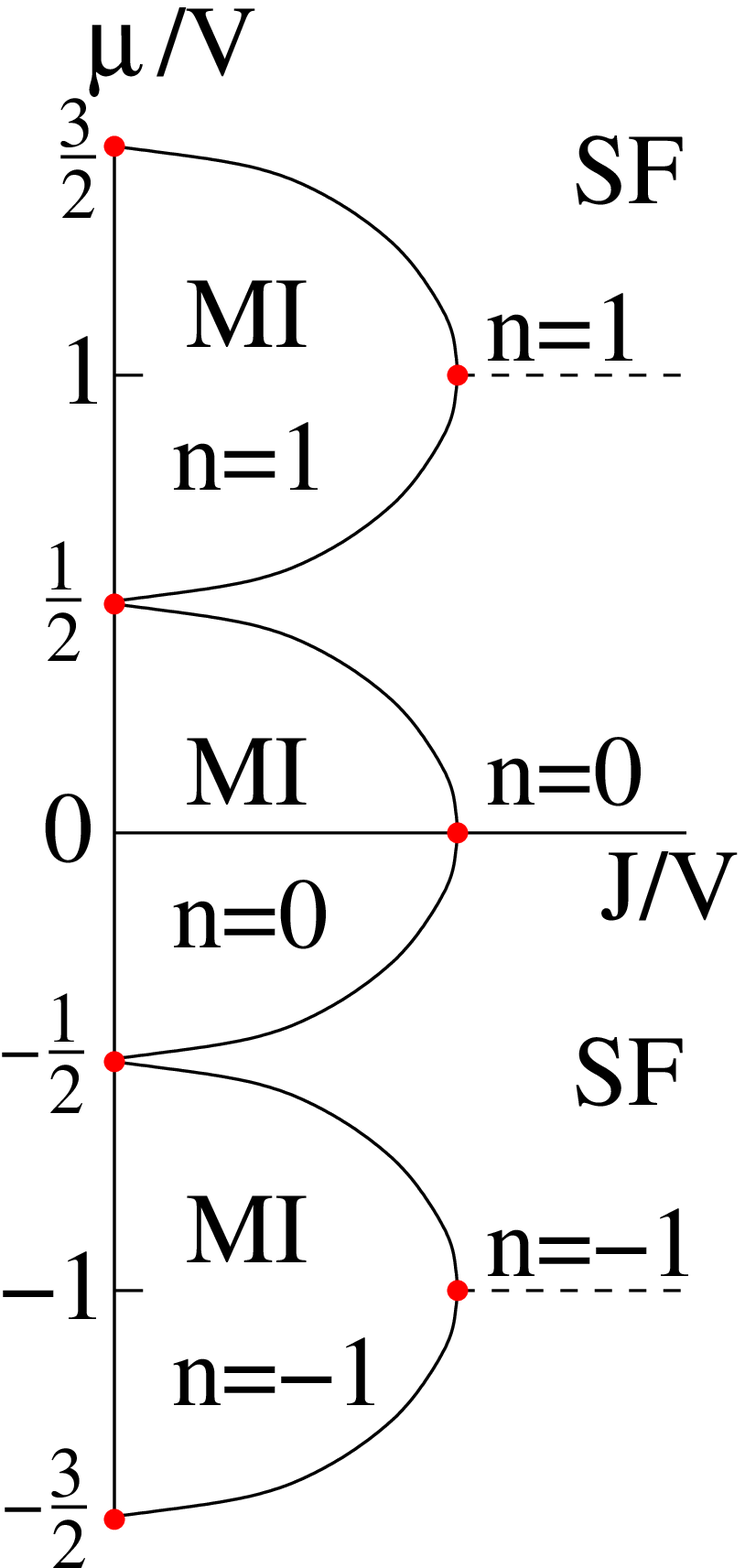,height=2.9in} \quad
\psfig{file=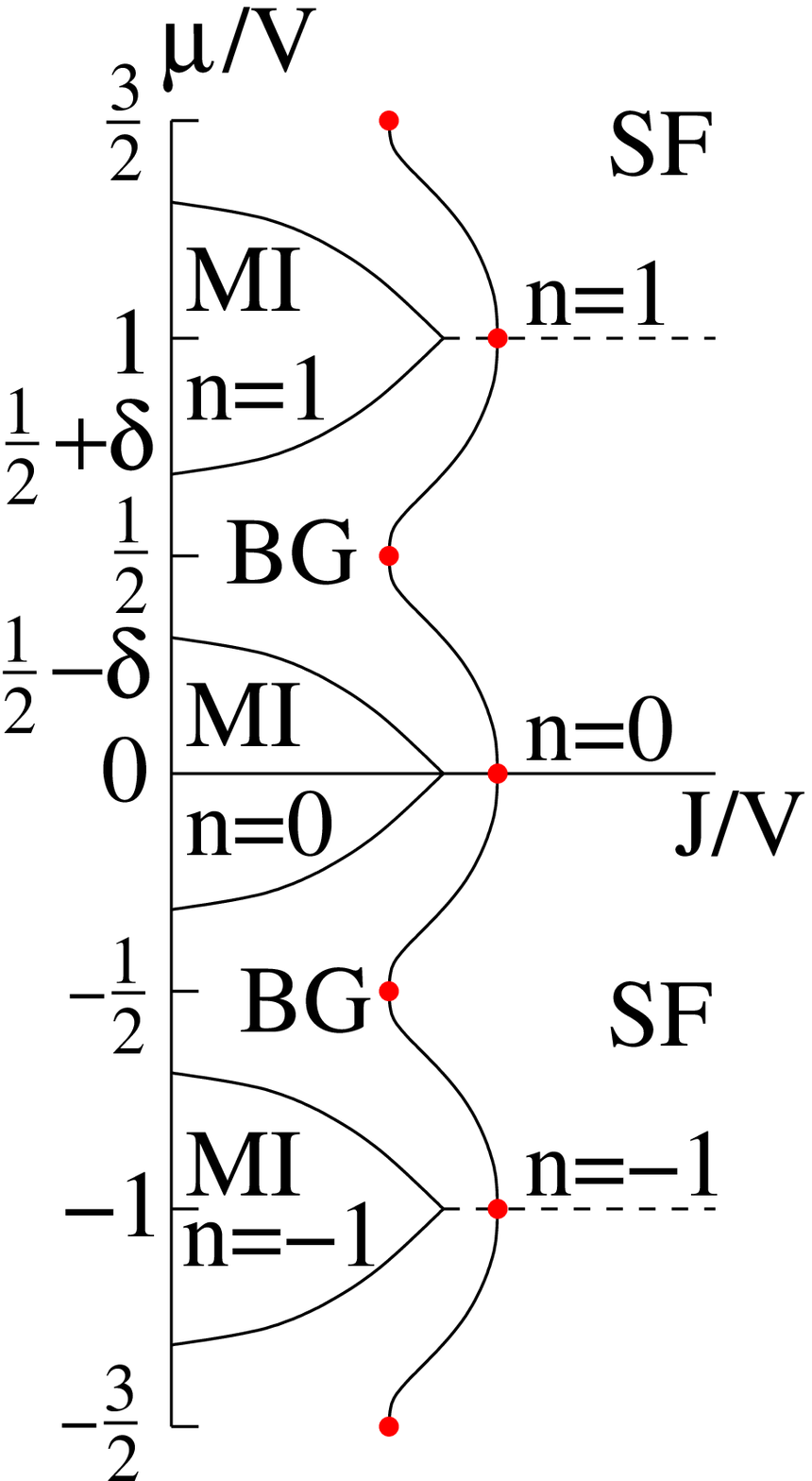,height=2.9in} \quad
\psfig{file=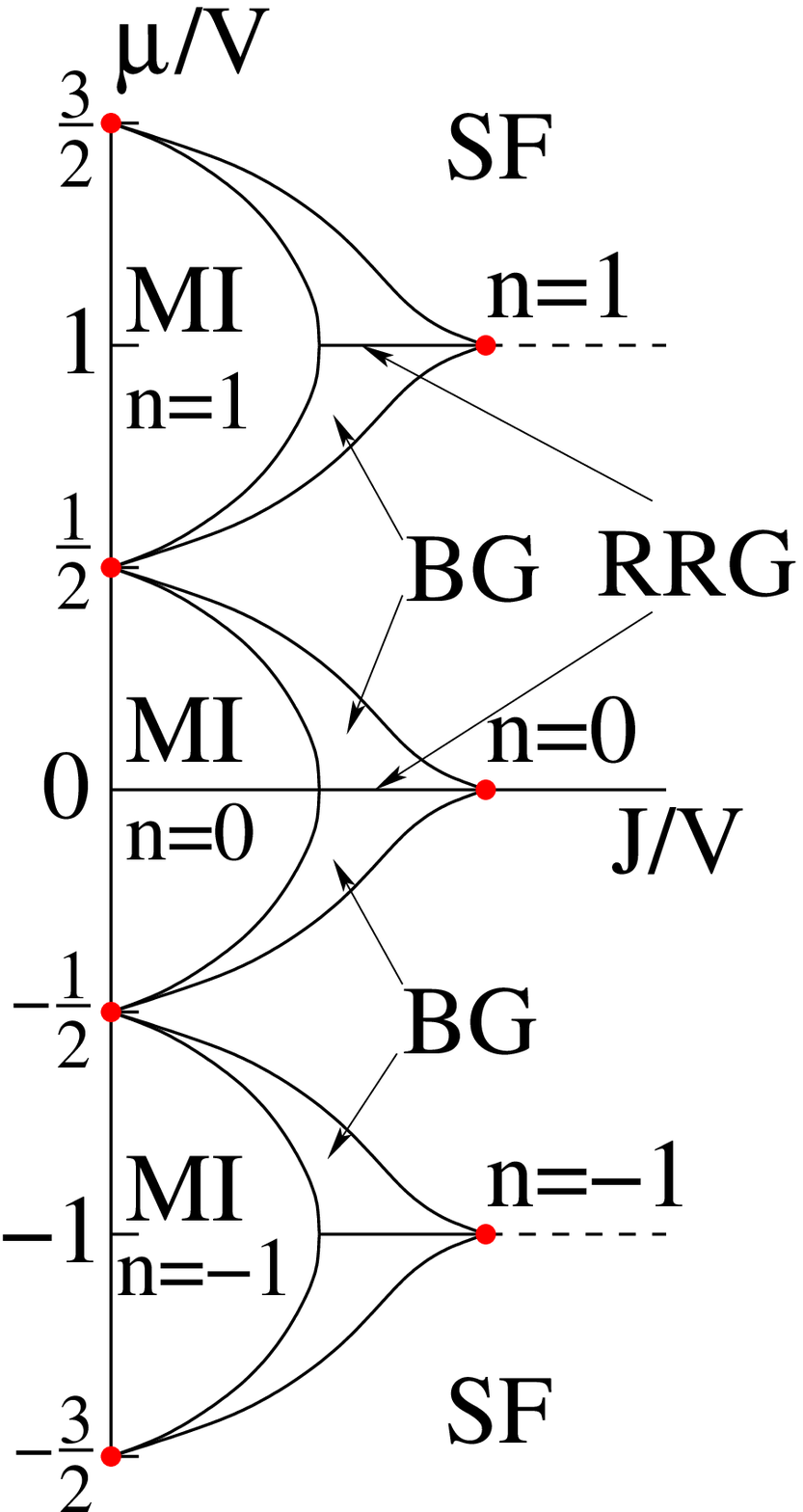,height=2.9in}}

\caption{Schematic phase diagrams for the Josephson junction model
(adapted from Ref.\ \protect\refcite{WM08}). Periodicity in $\mu/V$
is evident. Even site energy distributions have also been assumed,
hence the symmetry under $\mu \to -\mu$ as well. \textbf{Left:}
Clean case. \textbf{Center:} Generic site-disordered case, with
$\delta = \Delta/V$. \textbf{Right:} Case of pure hopping disorder.
The compressible Bose glass (BG) is replaced an incompressible
random rod glass (RRG) on the special lines at integer filling. The
penetration of the superfluid phase all the way to $J=0$ at
half-integer filling may be understood via a mapping to a
spin-$\frac{1}{2}$ quantum planar XY model with random exchange, but
vanishing out-of-plane magnetic field.\protect\cite{WM08}}

\label{fig:josephson_phases}
\end{figure}

\subsection{Phase diagrams for the Josephson junction model}
\label{sec:jjuncphases}

For comparison, phase diagrams for the Josephson junction model are
shown in Fig.\ \ref{fig:josephson_phases}. The clean and site
disordered models (left and center panels) are essentially identical
to their Bose-Hubbard counterparts (Figs.\ \ref{fig:hopping} and
\ref{fig:ranhopping}, respectively), except for the periodicity in
$\mu/V$, and symmetry under inversion $\mu \to -\mu$ for evenly
distributed $u_i$ (reflecting the existence of an exact, rather than
effective, particle-hole symmetry).

On the other hand the case of pure hopping disorder (right panel of
Fig.\ \ref{fig:josephson_phases}) has no Bose-Hubbard analogue. In
the absence of an exact particle-hole symmetry, random $J_{ij}$
always produces, in a renormalization group sense, an effective
random site energy even if none is present in the original
model---hence the choice of nonrandom $J$ in (\ref{2.1}). The exact
particle-hole symmetry for integer and half-integer $\mu/V$
drastically changes the character of the model along these lines.
For integer $\mu/V = n_0$, although the Mott gap closes at a finite
value $J = J_M$, the random rod glass (RRG) phase (see Sec.\
\ref{sec:pathint}) beyond it is incompressible. However, instead of
vanishing over a finite interval, for $J > J_M$ the compressibility
is finite for any $\mu/V \neq n_0$ (thus indicating a Bose glass
phase), and vanishes with an essential singularity, $\sim
e^{-c_0/|\mu/V-n_0|}$, as $|\mu/V - n_0| \to 0$.\cite{WM08} There
are corresponding stretched exponential tails, as opposed to the
Mott phase linear exponential tail, in the temporal correlation
function.\cite{WM08}

At half-integer $\mu/V$ the particle-hole symmetry has a different
effect, eliminating the glassy phase for all nonzero $J$. The model
may be mapped to a spin-$\frac{1}{2}$ quantum XY model, with the the
$u_i$ corresponding to mean-zero $z$-axis magnetic fields. If $u_i
\neq 0$, then for sufficiently small $J$ it is energetically
favorable for a large fraction of the spins to align with the field
along $z$, destroying the in-plane ferromagnetic order that
signifies superfluidity---this is the Bose glass phase. Only if $u_i
\equiv 0$ can the in-plane order, at least in higher dimensions $d >
1$,\footnote{For a strong-disorder approach to the 1D problem,
complementary to the weak disorder limit treated by
Kosterlitz-Thouless-type theories, see Refs.\
\refcite{Refaela,Refaelb}.} be maintained for arbitrarily small
exchange coupling.

\section{Droplet model of the Bose glass phase and the MI--BG phase
boundary}
\label{sec:droplet}

Before describing more formal theoretical approaches, I turn here to
some very simple arguments that provide a more intuitive picture of
the Bose glass phase, determine the exact vertical width of MI--BG
phase boundary (Fig.\ \ref{fig:mibg_bdy}), and establish essentially
rigorously that there can never be a direct MI--SF transition. As in
the clean case, for given $J,n_0$ the Mott phase is insensitive to
$\mu$, and described a unique ground state wavefunction.  The phase
boundary again occurs when $\mu$ coincides with with the lowest
lying single particle (or hole) excitation $\mu_\pm(J,\Delta,n_0) =
\pm[E(n_0 N_L \pm 1,\Delta,J)-E(n_0 N_L,\Delta,J)]$. The nature of
these excitations, and thereby the values of $\mu_\pm$, will now be
established.

\subsection{Low-lying particle and hole excitations}
\label{sec:excite}

Consider the addition of a single particle to the system. The energy
required can never be less than it would be if all site energies
were as small as possible, $u_i = -\Delta$ for all $i$. This
situation is equivalent to a simple shift, $\mu \to \mu-\Delta$, and
the bound $\mu_+(J,\Delta,n_0) \geq \mu_+(J,0,n_0) - \Delta$ follows
immediately. On the other hand, with an exponentially small density,
scaling as $p(R) \sim e^{-[R/R_0(\delta)]^3}$, where $R_0$ is some
length scale depending on the precise distribution of the $u_i$,
there will be rare, isolated ``droplets'' enclosing a sphere of
arbitrarily large radius $R$, on which all $u_i \leq -\Delta +
\delta u$ are nearly uniform, and within an arbitrarily small energy
$\delta u$ of $-\Delta$.  The single particle excitation energy of
such a droplet (above the uniform $u_i = -\Delta$ value) scales as
$\delta \varepsilon \approx c_1 \delta u + c_2 J (d/R)^2$, where
$c_1,c_2$ are appropriate geometrical constants, and the second term
represents the ground state of a particle in a box of radius
$R$.\footnote{More correctly, for larger $J$, one should replace $J
d^2$ in the second term of $\delta \varepsilon$ by a quantity
scaling with the background correlation length $\xi(J)$.} In an
infinite system, both terms can be made as small as one wishes, and
one obtains the opposite inequality $\mu_+(J,\Delta,n_0) \leq
\mu_+(J,0,n_0) - \Delta$. One therefore concludes that
$\mu_+(J,\Delta,n_0) = \mu_+(J,0,n_0) - \Delta$, and by an identical
argument for the hole excitations, that $\mu_-(J,\Delta,n_0) =
\mu_-(J,0,n_0) + \Delta$: the upper and lower pure system phase
boundaries each simply translate inwards by $\Delta$. This result is
sketched in Fig.\ \ref{fig:mibg_bdy}. The tip of the Mott lobe
occurs at the point $J = J_M(\Delta,n_0)$ where the two curves
intersect, i.e., $\mu_+(J_M,0,n_0)-\mu_-(J_M,0,n_0) = 2\Delta$,
thereby replacing the critical singularity at $J_c, \Delta = 0$ by a
simple slope discontinuity.

\begin{figure}[th]

\centerline{\psfig{file=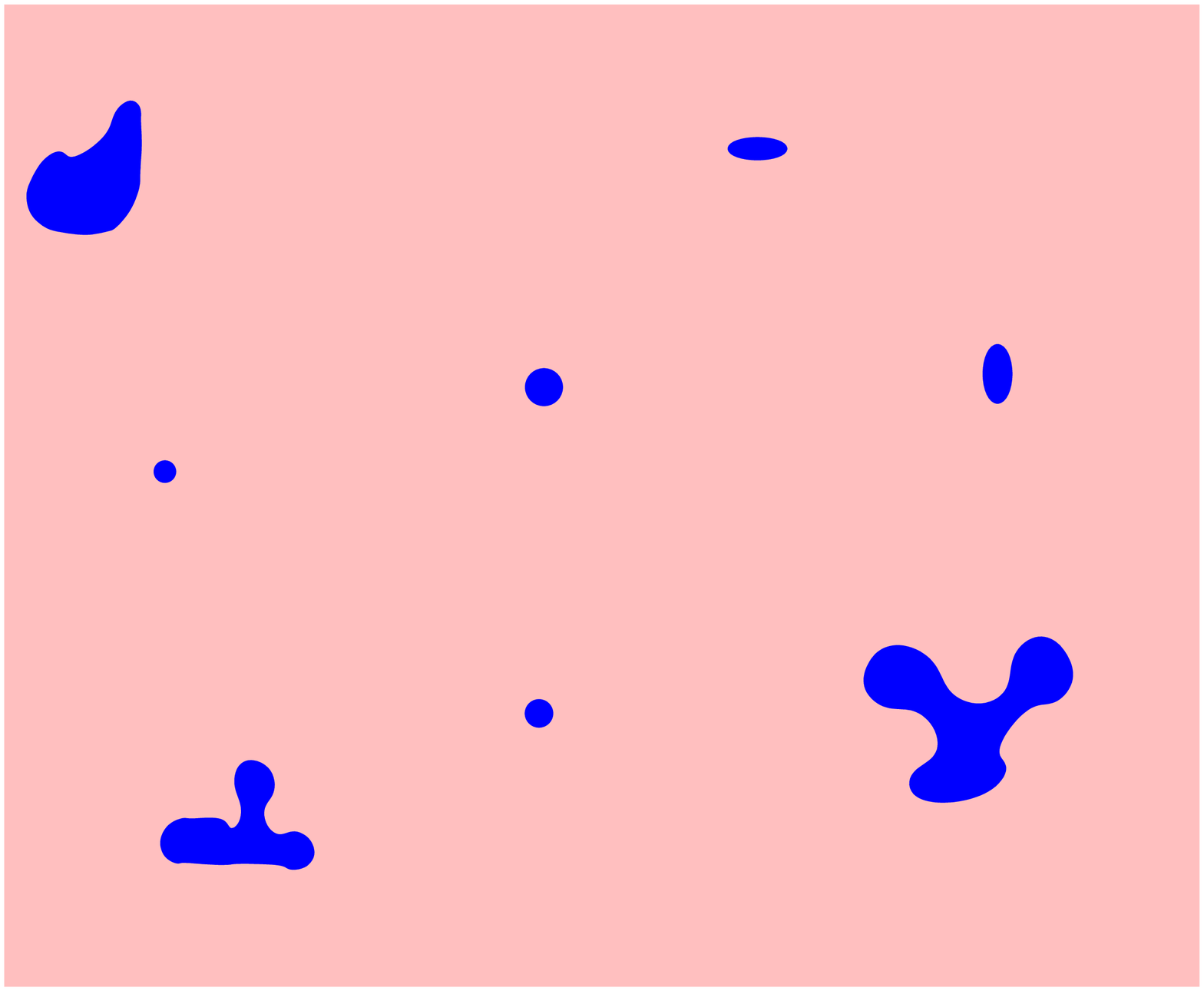,width=2.4in} \quad
\psfig{file=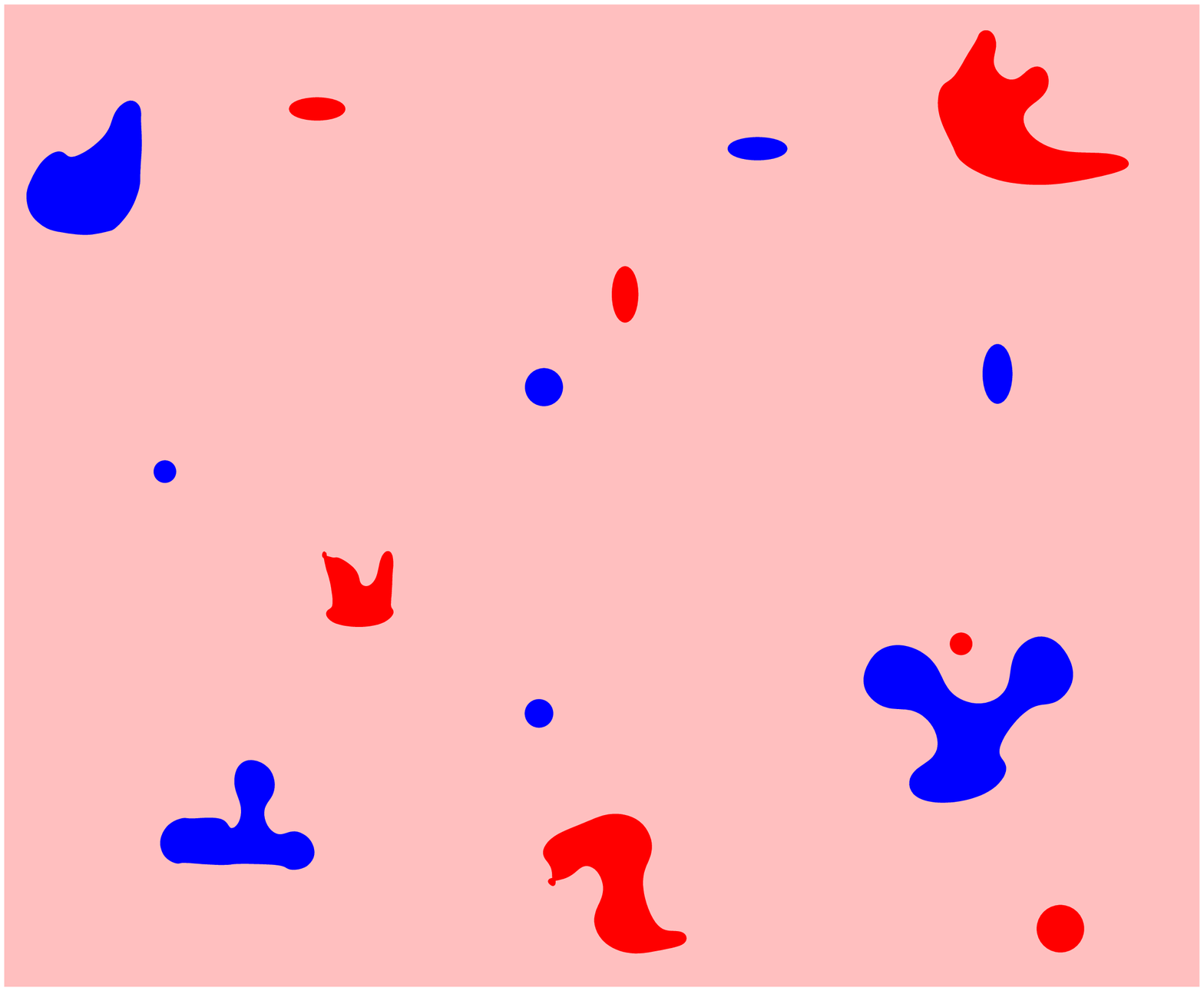,width=2.4in}}

\caption{Schematic illustration of the droplet model of the Bose
glass phase for the random site energy model (adapted from Ref.\
\protect\refcite{WM08}). \textbf{Left:} Exiting the Mott lobe at
fixed $J < J_M$ below its tip with increasing $\mu >
\mu_+(J,\Delta)$ [or decreasing $\mu < \mu_-(J,\Delta)$]. For
arbitrarily small $\epsilon = \mu - \mu_+(J,\Delta)$ there will be
an (exponentially small) density of droplets with volume large
enough that the energy gap to adding a particle is smaller than
$\epsilon$. \textbf{Right:} Exiting the Mott lobe from its tip at
fixed $\mu$ with increasing $J > J_M$. For arbitrarily small $\Delta
J = J - J_M(\Delta)$ there will similarly be arbitrarily large
droplets whose local chemical potential magnitude is above the
energy gap required to add a particle (previous droplets from the
left illustration), or below that required to remove a particle
(additional droplets).}

\label{fig:droplet}
\end{figure}

\begin{figure}[th]

\centerline{\psfig{file=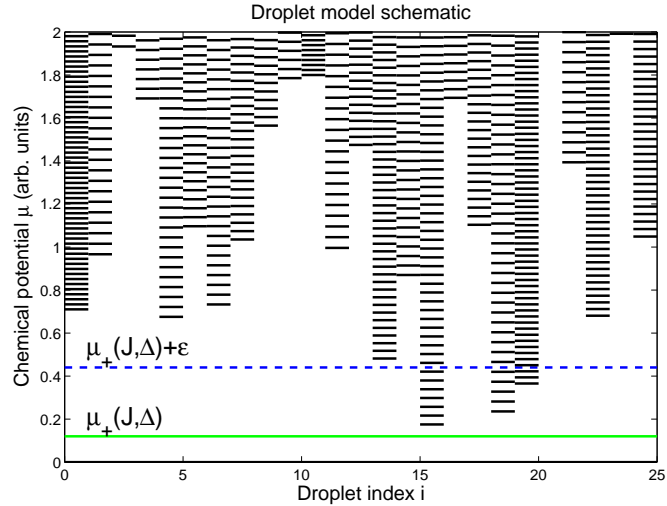,width=4.0in}}

\caption{Schematic illustration of spectrum of droplet excitations
(adapted from Ref.\ \protect\refcite{WM08}). Droplets are assumed to
be well separated, and independent (see Fig.\ \ref{fig:droplet}).
The horizontal axis is an arbitrary droplet index. The vertical axis
shows the critical chemical potential levels (in arbitrary units) at
which a new particle is added for each droplet. The lower bound of
the excitation spectrum lies at the Mott phase boundary
$\mu_+(J,\Delta) = \mu_+(J,0)-\Delta$ (solid horizontal line).  For
any $\epsilon \equiv \mu - \mu_+ > 0$ (dashed horizontal line),
there will be a finite (but exponentially small) density of large
droplets with local chemical potential lying sufficiently far above
$\mu_+$ that one or more extra particles are added. The total number
of particles added is given by counting the number of ``occupied''
levels below $\mu$. For a given droplet, while the lowest excitation
depends primarily on the local chemical potential, subsequent
particles are added in sequence, with gaps scaling as $1/\kappa_+ V$
where $V$ is the droplet volume, and $\kappa_+(J,\mu_+)$ is the bulk
compressibility of the pure (superfluid) phase just above the Mott
phase boundary. In an infinite system there are infinitely many
droplets, implying a continuous distribution of excitation energies,
and the bulk density will increase continuously with increasing
$\epsilon > 0$.}

\label{fig:mibg_excite}
\end{figure}

\subsection{Superfluid droplets and the Bose glass phase}
\label{sec:sfdrops}

Now, as $\mu$ increases above $\mu_+$ (or decreases below $\mu_-$),
these excitations begin to be populated with particles (or
holes)---the system is compressible. As sketched in Fig.\
\ref{fig:droplet}, the droplets now behave like uniform dilute
superfluid regions just above or below the clean system phase
boundary. As illustrated in somewhat more detail in Fig.\
\ref{fig:mibg_excite}, particles enter a given droplet at discrete
values of increasing $\mu$, with gaps controlled by the droplet
volume and the bulk compressibility of the clean system. Although
the droplets are superfluid, they are (exponentially in
$1/|\mu-\mu_\pm|$) widely separated. Global superfluid long range
order would require that the particles be able to tunnel coherently
through the intervening Mott phase, producing a finite (but, at
best, exponentially small) superfluid density. Such would indeed be
the case, for example, for identical droplets arranged periodically
in a homogeneous background. However, in the present case, the
tunneling particle also experiences a residual random potential due
to the $u_i$ (suitably renormalized by many body effects in the Mott
background), and the usual Anderson localization arguments imply
that phase coherence cannot be maintained.

A similar argument pertains if one exits the Mott lobe near its tip,
by increasing $J > J_M$ at fixed $\mu$. In this case there are
simultaneously present an exponentially dilute [in $1/(J-J_M)$] set
of both particle and hole superfluid droplets (right panel of Fig.\
\ref{fig:droplet}). By the same Anderson localization argument,
neither set of droplets can support global phase coherence.

In summary, one concludes that a new compressible, insulting Bose
glass phase completely surrounds the Mott lobe, and a direct MI--SF
transition is impossible. However, as $|\mu-\mu_\pm|$ (or $J-J_M$)
increases further, the superfluid droplets grow in size, and new
ones are created (as $\mu$ overcomes larger and larger local $u_i$
values). Eventually they percolate sufficiently that global phase
coherence is stabilized, and a BG--SF transition takes place. This
transition is presumably the ultimate zero temperature description
of helium in Vycor.

\begin{figure}

\centerline{\psfig{file=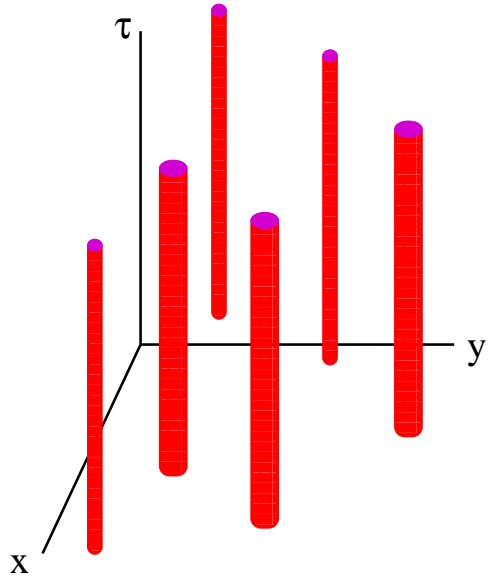,width=2.5in}}

\caption{Schematic illustration of random rod-like structure
generated by quenched disorder in quantum models (adapted from Ref.\
\protect\refcite{WM08}). The cylinders represent, for example,
random regions of enhanced or suppressed hopping strength or site
energies.}

\label{fig:rods}
\end{figure}

\section{Path integral formulations and universality classes}
\label{sec:pathint}

Given now a basic understanding of the underlying physics of the MI,
BG and SF phases, I turn now to more formal treatments that allow
one to clearly identify possible universality classes, and their
coarse-grained field theoretic descriptions.  One begins with the
path integral representation, based on the Trotter decomposition of
the partition function of the Josephson function model (\ref{2.2}),
$Z = \int D{\bm \phi} e^{-{\cal L}_J}$ with Lagrangian:\cite{WM08}
\begin{equation}
{\cal L}_J[{\bm \phi}] = \int_0^\beta d\tau
\left\{-\sum_{i,j} J_{ij} \cos[\phi_i(\tau)-\phi_j(\tau)]
+ K \sum_i [\partial_\tau \phi_i + i(u_i-\mu)]^2 \right\}
\label{5.1}
\end{equation}
in which $\beta = 1/k_B T \to \infty$, $K = 1/2V$, and $-\infty <
\phi_i(\tau) < \infty$ are continuous classical phase fields. As
usual, the quantum degrees of freedom give rise to an extra
(imaginary time) dimension $\tau$ in the effective classical model.
In general this extra dimension is highly anisotropic, and, except
in special cases, does \emph{not} simply lead to the same classical
model in one higher dimension. It is immediately evident, for
example, that the disordered quantities $J_{ij},u_i$ are
$\tau$-independent. The point like quenched disorder in the quantum
model therefore leads to a picture of \emph{columnar} or
\emph{rod-like} disorder in the classical model (Fig.\
\ref{fig:rods}).

In order to place (\ref{5.1}) in the context of more familiar field
theoretic models, we develop a coarse-grained, long wavelength
continuum approximation to (\ref{5.1}). Let $e^{i\phi_i(\tau)} \to
\psi({\bf x},\tau)$, and relax the sharp condition $|\psi| = 1$ on
the field magnitude in the standard fashion by adding terms
$\frac{1}{2} r_0 |\psi|^2 + \frac{1}{4} u_0 |\psi|^4$ to ${\cal L}$
(which constrains $|\psi| \approx \sqrt{-r_0/u_0}$). In addition the
cosine term is approximated by a squared gradient term $J |\nabla
\psi|^2$.  The result is an effective $\psi^4$ Lagrangian,
\begin{equation}
{\cal L}_c[{\bm \psi}] = \int_0^\beta d\tau \int d^dx
\left\{\frac{1}{2} |\nabla \psi|^2
-\frac{1}{2} \psi^*[\partial_\tau - g({\bf x})]^2 \psi
+ \frac{1}{2} r({\bf x}) |\psi|^2
+ \frac{1}{4} u_0 |\psi|^4 \right\}.
\label{5.2}
\end{equation}
Space and time have been rescaled so that the squared derivative
terms have unit coefficient. The random $J_{ij}$ is then subsumed
into a random $r({\bf x}) = r_0 + \delta r({\bf x})$, and $-\mu +
u_i$ has been subsumed into a random $g({\bf x}) = g_0 + \delta
g({\bf x})$, with $\langle \delta r \rangle = \langle \delta g
\rangle = 0$. Superfluidity corresponds to a nonzero anomalous
average $\langle \psi \rangle$, and one may think of $r_0$ as the
control parameter which takes one from the normal phase for large
positive $r_0$, to the superfluid phase for large negative $r_0$.

\begin{table}[pt]

{
\begin{tabular}{l}
Pure PH-sym [$(d+1)$-dimensional XY model]:
\\
\ \ ${\cal L}_0 = \int d^dx \int d\tau \left\{\frac{1}{2} |\nabla
\psi|^2 + \frac{1}{2} |\partial_\tau \psi|^2 + \frac{1}{2} r_0
|\psi|^2 + \frac{1}{4} u_0 |\psi|^4 \right\}$
\\
\\
Pure PH-asym [$d$-dimensional dilute Bose gas]:
\\
\ \ ${\cal L}_1 = \int d^dx \int d\tau \left\{\frac{1}{2} |\nabla
\psi|^2 - \frac{1}{2} \psi^*(\partial_\tau - g_0)^2 \psi +
\frac{1}{2} r_0 |\psi|^2 + \frac{1}{4} u_0 |\psi|^4 \right\}$
\\
\\
PH-sym RR [$(d+1)$-dimensional classical random rod model]:
\\
\ \ ${\cal L}_2 = \int d^dx \int d\tau \left\{\frac{1}{2} |\nabla
\psi|^2 + \frac{1}{2} |\partial_\tau \psi|^2 + \frac{1}{2} [r_0 +
\delta r({\bf x})] |\psi|^2 + \frac{1}{4} u_0 |\psi|^4 \right\}$
\\
\\
PH-asym RR [$(d+1)$-dimensional incommensurate random rod model]:
\\
\ \ ${\cal L}_3 = \int d^dx \int d\tau \left\{\frac{1}{2} |\nabla
\psi|^2 - \frac{1}{2} \psi^*(\partial_\tau - g_0)^2 \psi +
\frac{1}{2} [r_0 + \delta r({\bf x})] |\psi|^2 + \frac{1}{4} u_0
|\psi|^4 \right\}$
\\
\\
Statistical PH-sym [commensurate dirty boson problem]:
\\
\ \ ${\cal L}_4 = \int d^dx \int d\tau \left\{\frac{1}{2} |\nabla
\psi|^2 - \frac{1}{2} \psi^* [\partial_\tau - \delta g({\bf x})]^2
\psi + \frac{1}{2} [r_0 + \delta r({\bf x})] |\psi|^2 + \frac{1}{4}
u_0 |\psi|^4 \right\}$
\\
\\
Generic PH-asym [incommensurate dirty boson problem]:
\\
\ \ ${\cal L}_5 = \int d^dx \int d\tau \left\{\frac{1}{2} |\nabla
\psi|^2 - \frac{1}{2} \psi^* [\partial_\tau - g_0 - \delta g({\bf
x})]^2 \psi + \frac{1}{2} [r_0 + \delta r({\bf x})] |\psi|^2
\right.$
\\
\hskip1.25in $+\ \left. \frac{1}{4} u_0 |\psi|^4 \right\}$
\end{tabular}}

\caption{Continuum $\psi^4$ representation of models with various
types of disorder and various degrees of particle-hole symmetry. The
coefficients of $|\nabla \psi|^2$ and $|\partial_\tau \psi|^2$ have
been normalized to $\frac{1}{2}$. The control parameters $r_0$ and
$g_0$ are analogous to $J_0$ and $\mu$, respectively. Disorder in
the hopping strengths is represented by $\delta r$, while that in
the site energies is represented by $\delta g$. Both are independent
of $\tau$, and in field theoretic treatments, are taken as quenched
Gaussian random fields with zero mean and delta-function
correlations characterized by variances $\Delta_r$ and $\Delta_g$,
respectively. Disorder in the other parameters (including the unit
gradient-squared coefficients) may also be introduced, but produces
no new critical behavior.}

\label{table1}
\end{table}

One may now consider the various possible classes of phase
transition depending on the various terms, and their underlying
symmetries, that one keeps in (\ref{5.2}). Table \ref{table1} lists
the relevant special cases.

\subsection{Particle-hole symmetric transitions}
\label{sec:phsym}

The particle-hole symmetric model $\mu = 0$, $u_i \equiv 0$
corresponds to $g({\bf x}) \equiv 0$.  The Lagrangian ${\cal L}_0$
(clean system) or ${\cal L}_2$ (system with random hopping) is then
purely real, and $e^{-{\cal L}_{0,2}}$ may be interpreted in terms
of a classical probability density for the two-component field
$(\mathrm{Re}(\psi),\mathrm{Im}(\psi)]$. The squared-gradient
interaction in both space and time implies that the model is the
usual Landau-Ginzburg-Wilson representation of a ferromagnetic
XY-model in $(d+1)$-dimensions. In the clean case, time and space
are fully symmetric, and the transition (through the tips of the
Mott lobes in Fig.\ \ref{fig:hopping}) is in the universality class
of the usual classical $(d+1)$-dimensional XY model. With random
hopping, $\delta r({\bf x}) \neq 0$, strong anisotropy is generated
by the columnar disorder (see Fig.\ \ref{fig:rods}), and the
transition is in the universality class of the so-called
\emph{classical random rod problem.}\cite{D1980,BC1983,LP1984}

\subsection{Particle-hole asymmetric transitions}
\label{sec:phasym}

For nonzero $\mu$ or $u_i$, hence $g({\bf x})$, the Lagrangian is
complex, and a classical probabilistic interpretation is no longer
possible. The leading interaction is now a (purely imaginary) linear
time derivative $g \psi^* \partial_\tau \psi$, which generates an an
even stronger space-time anisotropy. For the clean system, $g_0 \neq
0$ but $\delta r({\bf x}) = \delta g({\bf x}) \equiv 0$ (the
Lagrangian ${\cal L}_1$), this term generates critical behavior
equivalent to the onset of superfluidity (described quantitatively
by the Bogoliubov model\footnote{See, e.g., Ref.\ \refcite{FW}. For
a more modern view, see Ref.\ \refcite{W1989}.} at zero density in a
dilute Bose gas.\footnote{The usual coherent state functional
integral formulation of the boson model (\ref{2.1}) corresponds to
an \emph{absent} $|\partial_\tau \psi|^2$ term, and a \emph{unit
coefficient} $\psi^*\partial_\tau \psi$ term, so that the model can
never possess an exact particle-hole symmetry. Treating both
particle-hole symmetric and asymmetric models requires the expanded
space of Lagrangians corresponding to (\ref{5.2}).}

If $g_0 = 0$ and $\delta g({\bf x}) \neq 0$ has an even
distribution, then the corresponding Lagrangian ${\cal L}_4$ has a
\emph{statistical particle-hole symmetry}---the $g \psi^*
\partial_\tau \psi$ term is locally nonzero, but zero on average
(the manifestation of this symmetry is more evident in the
replicated, disorder averaged Lagrangian discussed in Sec.\
\ref{sec:rg}). It will be argued below that the transition in this
case is identical to that for the generic Lagrangian ${\cal L}_5$ in
which all terms are nonzero, so that the statistical symmetry is
\emph{restored} at the critical point.\footnote{This idea has been
used to explain the vanishing of the Hall conductivity at magnetic
field-tuned superconducting transitions: Refs.\
\refcite{PHSa,PHSb}.} Intuitively, the expanding superfluid droplets
in the Bose glass phase as the superfluid phase boundary is
approached lead to a decoupling of the particle number from the
lattice potential. The discrete droplet excitation spectrum pictured
in Fig.\ \ref{fig:mibg_excite}, which replaces the individual
lattice site occupation spectrum, becomes denser and denser, the
statistics of the energy spectrum loses its up-down asymmetry, and
there is less and less distinction between the statistics of
particle and hole excitations as $\mu$ is varied.

Less surprisingly, the particle-hole asymmetric random rod model,
${\cal L}_3$, in which $g_0, \delta r({\bf x}) \neq 0$ but $\delta
g({\bf x}) \equiv 0$, is equivalent to the generic model. In the
renormalization group sense, the combined effect of $g_0$ and
$\delta r({\bf x})$ produces nonzero $\delta g({\bf x})$.
Intuitively, in the original boson model (\ref{2.1}), random hopping
produces regions of varying compressibility, and a uniform $\mu$
will produce a nonuniform density. As far as critical behavior goes,
this is indistinguishable from density fluctuations produced a
random site potential (Sec.\ \ref{sec:droplet} and Figs.\
\ref{fig:droplet}, \ref{fig:mibg_excite}).

\section{Quantum scaling theory}
\label{sec:scaling}

In describing the Vycor data in Sec.\ \ref{sec:intro} several zero
temperature critical exponents were introduced. Here the general
scaling theory of a QPT (which is not initially limited to the dirty
boson problem) is summarized, these exponents are placed in a more
general context, and scaling relations between them and more
familiar exponents are derived.\cite{FWGF}

Let $\delta$ be the thermodynamic control parameter (e.g., $J-J_c$,
$\mu-\mu_c$, or $r_0-r_{0,c}$), with critical point at $\delta = 0$.
The divergence of the spatial and temporal correlation lengths, $\xi
\approx \xi_0 |\delta|^{-\nu}$, $\xi_\tau \approx \xi_{0,\tau}
|\delta|^{-\nu_\tau}$ define critical exponents $\nu, \nu_\tau$. The
\emph{dynamical exponent} $z = \nu_\tau/\nu$ quantifies the
space-time anisotropy. In particular, the isotropic Lagrangian
${\cal L}_0$ must lead to $z=1$, while for the remaining models one
expects $z \neq 1$. The singular part of the free energy scales in
the form $F_s \approx A |\delta|^{2-\alpha}$, which defines the
analogue of a zero temperature ``specific heat'' exponent $\alpha$.
The \emph{quantum hyperscaling} relation relation $2-\alpha =
(d+z)\nu$ follows from the assumption that $F_s$, being an energy
density, scale also inversely with the space-time correlation volume
$(\xi_\tau \xi^d)^{-1} \approx (\xi_{\tau,0} \xi_0^d)^{-1}
|\delta|^{(d+z)\nu}$.

The critical behavior of the superfluid density, $\rho_s \approx
\rho_{s,0} |\delta|^\upsilon$ defines and exponent $\upsilon$. The
superfluid density is a torsional modulus quantifying the stiffness
of the order parameter, and is therefore derived from the free
energy increment $\Delta F_s = \frac{1}{2} \rho_s v_s^2$ (which must
be part of $F_s$ because $\rho_s$ vanishes in the insulating phase,
and is therefore a singular quantity), where the superfluid velocity
${\bf v}_s = \hbar {\bf k}_s/m$ is generated by a long wavelength
gradient ${\bf k}_s = \nabla \phi$ in the order parameter phase
(corresponding to a helical twist in the order parameter itself).
Since ${\bf k}_s$ is an inverse length, it must scale as $\xi^{-1}$,
and  one obtains the \emph{quantum Josephson relation} $\upsilon =
2-\alpha - 2\nu = (d+z-2)\nu$.  If, as in the Vycor experiments, one
uses the particle density difference as the control parameter,
$\delta = \rho - \rho_c$, then the exponent $\omega$ defined in
Fig.\ \ref{fig:hevycor} is given simply by $\omega = \upsilon$.

The temperature is similarly included via the temporal finite size
scaling form $F_s \approx A |\delta|^{2-\alpha} Y(\beta/\xi_\tau)$,
in which $Y(y)$ is a universal scaling function with $Y(\infty) =
1$. For finite $\beta$ the argument is finite, and $Y(y)$ must
interpolate between the zero and finite temperature critical
behaviors. In particular, the finite temperature transition must
occur at some value $y = y_c$, which leads to the relation $k_B
T_c(\delta) \approx (y_c \xi_{\tau,0})^{-1} |\delta|^{z\nu}$.
Referring again to Fig.\ \ref{fig:hevycor}, one therefore obtains
$\rho_s(T=0) \sim T_c^\theta$ with $\theta = (d+z-2)/z$. One obtains
therefore the ratio $\omega/\theta = z\nu = \nu_\tau$.

Using the Vycor data experimental values $\omega = 1.7 \pm 0.3$ and
$\theta = 1.25 \pm 0.2$ in $d=3$, one obtains the reasonable result
$\nu_\tau = 1.4 \pm 0.3$. However, the implied relations $z =
1/(\theta-1)$ and $\nu = \omega/\theta z$ produce the remarkably
unsatisfactory bounds $2.2 \leq z \leq 20$ and $0.07 \leq \nu \leq
0.64$. Unfortunately, more accurate bounds would require much higher
quality data, but even with the intervention of 25 years, the
original experiments\cite{Reppy1983} would be extremely difficult to
improve upon.

\subsection{Compressibility and $z$ vs.\ $d$}
\label{sec:zvsd}

If one uses $\delta = \mu-\mu_c$, the compressibility $\kappa =
\partial n/\partial \mu = -\partial^2 F/\partial \delta^2$ has a
singular part $\kappa_s \sim |\delta|^{-\alpha}$ controlled by the
zero temperature specific heat exponent (this is true more generally
so long as variation in $\delta$ leads to variation in the density).
However, the \emph{total} compressibility also has an interpretation
as a \emph{temporal superfluid density:} the free energy increment
due to temporal phase twists $\omega_s = \partial_\tau \phi$ is
given by $\Delta F_\tau = \frac{1}{2} \kappa \omega_s^2$. This
identification follows from (\ref{5.1}) since the imposition of the
phase twist $\omega_s$ is equivalent to the shift $\mu \to \mu +
i\omega_s$, and it follows that $\partial^2 \Delta F_\tau/\partial
\omega_s^2 = -\partial^2 F/\partial \mu^2 = \kappa$.  By analogy to
$\rho_s$, if one assumes that $\Delta F_\tau$ is part of $F_s$, and
proposes that $\omega_s$ scales as $\xi_\tau^{-1}$, then one obtains
$\kappa \approx \kappa_0 |\delta|^{\upsilon_\tau}$ with
$\upsilon_\tau = (d-z)\nu$. The physical fact that $\kappa$ is
finite and nonzero through the transition (with $\kappa_s$ yielding
a small correction since one expects $\alpha < 0$) then leads to the
proposed scaling relation $z=d$.\cite{FWGF}

However, this argument lies on very shaky ground because it is
difficult to reconcile $\kappa_s$ and $\kappa$ obeying separate
critical scaling relations. In fact, the difference
$\kappa-\kappa_s$, being finite through the BG--SF transition, is
dominated by the analytic background, and $\Delta F_\tau$ should be
likewise.\cite{WM07} Moreover, since the shift $\mu \to \mu +
i\omega_s$ simply leads to a small adjustment of a parameter that is
already nonzero, the scaling assumption $\omega_s \sim
\xi_\tau^{-1}$ is inappropriate. \emph{Only if} $\mu = 0$, $u_i
\equiv 0$, so that $\omega_s$ breaks particle-hole symmetry, should
one expect such a scaling.\footnote{For $\rho_s$, the spatial phase
twist corresponds to the shift $\nabla^2 \to (\nabla + i{\bf
k}_s)^2$, introducing a momentum term $i{\bf k}_s \cdot \psi^*
\nabla \psi$ which breaks a spatial inversion symmetry. It is for
this reason that one expects the critical scaling of $k_s$ with
$\xi^{-1}$.} Indeed, for the random rod problem, $\kappa = \kappa_s$
is entirely singular since, like $\rho_s$, it vanishes identically
in the RRG phase. Consistently, one finds $\upsilon_\tau > 0$, hence
$z < d$, at that transition.\cite{D1980,BC1983,LP1984} Consistently
as well, the control parameter $\delta = J-J_c$ in that case does
not couple to the density, and $\kappa$ is no longer related to the
exponent $\alpha$.

In conclusion, at both the RR--SF and BG--SF transitions $z$ is
expected to remain an independent exponent, undetermined by any
simple scaling relation. Recent high resolution quantum Monte Carlo
simulations in $d=2$ support this view, finding $z = 1.40 \pm 0.02$
at the BG--SF transition.\cite{Baranger06}

\subsection{Universal critical sheet conductance in $d=2$}
\label{sec:univcond}

The conductivity is related to the superfluid density via the zero
frequency limit of $\sigma = \rho_s/i\omega$. If one assumes that
$\omega$, being an inverse time, scales as $\xi_\tau^{-1}$ (this is
reasonable here, because $\omega$ is a real dynamic frequency, not
an imposed phase twist, and $\sigma$ is indeed singular), one
obtains $\sigma \approx \sigma_0 |\delta|^\Sigma$ with $\Sigma =
2-\alpha - (z+2) \nu = (d-2)\nu$. This has the remarkable
implication that, very generally, in $d=2$ the conductivity, while
vanishing in the insulating phase, and diverging in the superfluid
phase, has a \emph{finite constant} value $\sigma_c$ right at the
critical point. Experiments on disordered superconducting thin
films, where $\delta$ is the film thickness or an applied magnetic
field, provide some support for this\cite{HLG1989,HP1990,F1990} (see
also Ref.\ \refcite{GM1998} and references therein). Moreover,
\emph{quantum hyperuniversality},\cite{KW91} which predicts not only
that the critical exponents for $F_s$ and $(\xi_\tau \xi^d)^{-1}$
coincide, but also that the amplitude combination $F_s \xi_\tau
\xi^d \approx A \xi_{\tau,0} \xi_0^d$ also be a universal number,
predicts that $\sigma_c = A \xi_{\tau,0} \xi_0^2$ coincides with
this amplitude, and therefore should be
\emph{universal}.\footnote{The scaling theory is general, and does
not require any identification between the dirty superconducting and
superfluid systems. However, in modeling thin films close to the
transition, it is typically argued that Cooper pairs may indeed by
treated as bosons, hence that the transition lies in the same
universality class as the dirty boson problem, and $\sigma_c$ may be
computed within this model.  See, e.g., Ref.\ \refcite{SWGY1992}.}
This has much less experimental support---a large range of estimated
$\sigma_c$ values are found.\cite{HLG1989,HP1990,GM1998} On the
other hand, critical amplitude ratios are far more difficult to
measure accurately than critical exponents, and higher quality data
would again be desirable.

\section{Renormalization group approaches}
\label{sec:rg}

The dirty boson problem has so far discussed from a phenomenological
point of view, emphasizing elementary excitations, symmetry
principles, and results that follow from general scaling relations.
This review is concluded with a brief survey of renormalization
group approaches to quantitative evaluation of exponents and other
quantities.

\subsection{One-dimensional models}
\label{sec:1d}

There is one limit where an exact solutions exist, namely
$d=1$.\cite{GS87,GS88,FWGF,Refaela,Refaelb}  The model maps to a
classical 2D fluctuating sine-Gordon-type interface roughening
model:\cite{WM08}
\begin{equation}
{\cal L}_\mathrm{SG}[{\bf h}] = \int dx \int d\tau
\left\{\frac{1}{2} K(x) (\partial_\tau h)^2
+ \frac{1}{2} V (\partial_x h)^2
- m(x) \partial_x h - 2 y_0 \cos(2\pi h) \right\},
\label{7.1}
\end{equation}
where $h(x,\tau)$ is the interface height at a given space-time
point.\footnote{An extension to $d = 1+\epsilon$ dimensions has also
been proposed, based on inserting the engineering dimensions
appropriate to higher $d$ into the one-dimensional RG flows: see
Ref.\ \refcite{H1998}. However, in the absence of a form for the
surface roughening type Hamiltonian generalizing (\ref{7.1}) for
noninteger $d > 1$ (the particle-vortex duality transformations used
to obtain such models requires integer $d$), there is presently no
rigorous support for this approach.} The hopping disorder leads to
disorder in the interface tension coefficient $K(x)$, while the
chemical potential and site disorder leads to the disordered tilt
potential $m(x) = m_0 + \delta m(x)$. The cosine term prefers
integer values of $h$, a reflection of the original integer boson
site occupancy. At sufficiently large $y_0$ this term pins the
interface at a fixed integer height, with small fluctuations about
it. This is the Mott insulating phase. For the 1D random rod
problem, $m(x) \equiv 0$, at small $y_0$ the fluctuations are able
to unpin the interface, leading to logarithmically divergent large
scale fluctuations: this corresponds to the superfluid phase. The
disorder in $K$ turns out to be irrelevant at this transition, so
both the clean and random rod problems are described by the usual
classical 2D Kosterlitz-Thouless theory, with, in particular $z=1$
and critical correlation decay exponent $\eta = \frac{1}{4}$ (see,
e.g., Ref.\ \refcite{JKKN}).

On the other hand, nonzero $m(x)$ couples directly to the interface
slope, and can have a much stronger effect. At a sufficiently large
average value, $m_0$, there is a transition to an interface with an
average constant spatial tilt. The value of $m_0$ at the onset of
this tilt corresponds to the Mott excitation gap. For sufficiently
small $K,V$ the fluctuations about the mean tilt completely wash out
the cosine term and are again logarithmically divergent on large
scales: this is again the superfluid phase. However, at intermediate
values there is enough residual effect of the cosine term to
constrain the interface to finite fluctuations about this tilted
state---this corresponds to the Bose glass phase. The BG--SF
transition is described by a modified Kosterlitz-Thouless
theory\footnote{In the replicated model, the tilt disorder replaces
the usual $cos(2\pi h)$ term by a replica interaction term of the
form $\cos\{[2\pi[h_\alpha(x,\tau)-h_\beta(x,\tau')]\}$. This term
generates renormalization group flows similar to those of classical
Kosterlitz-Thouless (Ref.\ \refcite{JKKN}) but with sufficiently
different geometry that different universal critical properties are
produced.} which produces different universal critical
properties.\cite{GS87,GS88,FWGF} For example, although $z=1$ still,
one now finds $\eta = \frac{1}{3}$. Moreover, near the transition
one may absorb $m_0$ into a shift $h \to h - (m_0/V) x$, so that
only $\delta m$ enters. This explicitly demonstrates the asymptotic
restoration of statistical particle-hole symmetry in the 1D model.

\subsection{Epsilon-expansions in higher dimensions}
\label{sec:epsilon}

We now turn to epsilon-expansion type approaches in higher
dimensions that have proven so useful in quantifying classical
critical phenomena. Renormalization group approaches are generally
based on ``replicated'' Lagrangians, in which the disorder has been
integrated out at the expense of introducing $p$ copies,
$\psi_\alpha$, $\alpha = 1,2,\ldots,p$ of the original field that
all interact with the same quenched disorder $\delta r,\delta g$,
but are otherwise independent, with the formal limit $p \to 0$ taken
at the end. Using the standard delta-correlated Gaussian disorder
model, with variances $\Delta_r$ and $\Delta_g$, the replicated
$\psi^4$ Lagrangian takes the form ${\cal L}_C^{(p)} = {\cal
L}_{C,1}^{(p)} + {\cal L}_{C,2}^{(p)}$ with,\cite{WM08,MW96}
\begin{eqnarray}
{\cal L}_{C,1}^{(p)} &=& \sum_{\alpha = 1}^p
\int_0^\beta d\tau \int d^dx
\left\{\frac{1}{2} |\nabla \psi_\alpha|^2
-\frac{1}{2} \psi_\alpha^* (\partial_\tau - g_0)^2 \psi_\alpha
+ \frac{1}{2} r_0 |\psi_\alpha|^2
+ \frac{1}{4} u_0 |\psi_\alpha|^4 \right\}
\nonumber \\
{\cal L}_{C,2}^{(p)} &=& -\frac{1}{2} \sum_{\alpha,\beta = 1}^p
\int_0^\beta d\tau \int_0^\beta d\tau' \int d^dx
\left\{\Delta_r |\psi_\beta({\bf x},\tau)|^2
|\psi_\alpha({\bf x},\tau')|^2 \right.
\nonumber \\
&&+\ \left. \Delta_g [\psi_\alpha^* \partial_\tau \psi_\alpha
- g_0 |\psi_\alpha|^2]({\bf x},\tau)
[\psi_\beta^* \partial_\tau \psi_\beta
- g_0 |\psi_\beta|^2]({\bf x},\tau') \right\}.
\label{7.2}
\end{eqnarray}
The disorder average leads to quadratic interactions between the
replicas that are of infinite range in $\tau$, a reflection of its
columnar character, and this is what generates new critical
behavior. The random rod model corresponds to $g_0=0$, $\Delta_g =
0$, so that the Lagrangian is separately invariant under reversal of
$\tau$ and $\tau'$. The case of statistical particle-hole symmetry
corresponds to $g_0=0$, $\Delta_g > 0$, which leads to a symmetry
only under \emph{simultaneous} reversal of $\tau,\tau'$.

Now, imagine beginning with the random rod model, and perturbing it
in various ways. Nonzero $\Delta_g$ leads, in a heuristic Fourier
space notation, to a $\Delta_g \omega^2 \psi^4$ correction to the
$\Delta_r \psi^4$ term already present.  Under naive power counting,
since critical fluctuations are dominated by large scales and long
times, increasing powers of frequency lead to less important terms,
and one would expect the former term to be strongly irrelevant
relative to the latter, i.e., that symmetric site disorder should
\emph{not} destabilize the RR critical behavior.  On the other hand,
$g_0$ introduces a $i g_0 \omega \psi^2$ perturbation, which by
power counting dominates the $\omega^2 \psi^2$ term, and \emph{is}
expected to lead to new critical behavior. This is certainly
confirmed in the clean model, where it generates the crossover from
$(d+1)$-dimensional XY behavior to weakly interacting Bose gas
behavior (Sec.\ \ref{sec:excite}). However, the notion of asymptotic
restoration of particle-hole symmetry requires that, in the
disordered case, this term ultimately be subdominant to the
$\Delta_g$ term.

\begin{figure}[th]

\centerline{\psfig{file=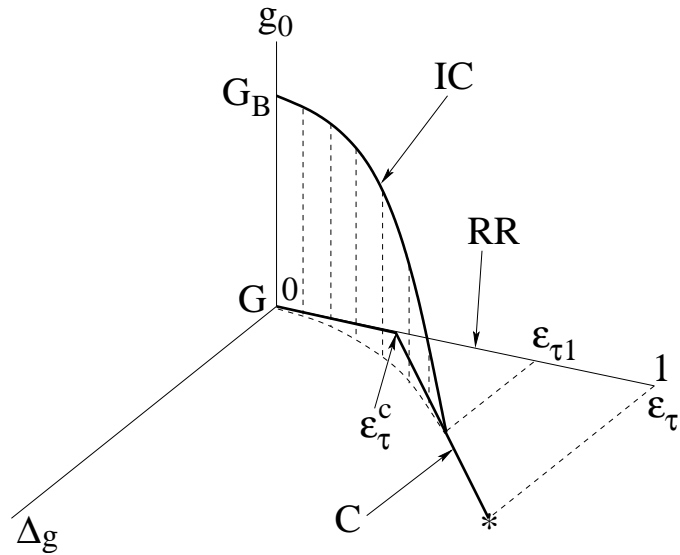,width=3.5in}}

\caption{Proposed behavior of the random rod (RR), commensurate
(statistically particle-hole symmetric) dirty boson (C), and
incommensurate (fully particle-hole asymmetric) dirty boson (IC)
fixed points as functions of $\epsilon_\tau$ (adapted from Refs.\
\protect\refcite{WM08,MW96}). The $\Delta_r$ axis has been
suppressed. Here $G$ and $G_B$ are commensurate and incommensurate
Gaussian fixed points. The commensurate fixed point bifurcates away
from the random rod fixed point at $\epsilon_\tau^c \simeq
\frac{8}{29}$. At $\epsilon_\tau = \epsilon_{\tau 1} \simeq
\frac{2}{3}$, C and IC merge and at $\epsilon_\tau = 1$, C is the
stable fixed point that describes the physical dirty boson problem.}

\label{fig:fixedpoints}
\end{figure}

These paradoxes find their resolution in the fact that naive power
counting is exact only at Gaussian critical points, and
approximately valid only when interactions (terms of order higher
than $\psi^2$) play a small role. If their role is large, then
different terms can indeed exchange dominance.  In order to explore
such notions quantitatively, it is convenient to have a model
parameter that, when varied, interpolates between the near-Gaussian
and strongly non-Gaussian limits. In classical critical phenomena,
this parameter is $\epsilon = 4-d$: the standard epsilon expansion
is based on the fact that the critical behavior is Gaussian for $d
\geq 4$, and nearly so for small $\epsilon > 0$. The corresponding
approach for problems with columnar disorder requires not only that
$d$ be close to four, but also that the dimension $\epsilon_\tau$
\emph{of the columns} be small---essentially this is an expansion
about the point-disorder limit. With $D = d + \epsilon_\tau$ being
the total dimensionality, and $\epsilon = 4-D$, one may develop
renormalization group recursion relations for the random rod problem
which have a perturbatively accessible critical fixed point when
$\epsilon,\epsilon_\tau$ are both small.\cite{D1980,BC1983,LP1984}
The physical problem in $d=3$ is recovered for $\epsilon_\tau = 1$
and $\epsilon = 0$.

This double epsilon-expansion is very poorly behaved---with
coefficients having very strong dependence on these parameters, and
higher order terms dominating lower order ones unless
$\epsilon,\epsilon_\tau$ are extremely small. Nevertheless, as shown
in Fig.\ \ref{fig:fixedpoints}, a very compelling picture
emerges.\cite{MW96} For sufficiently small $\epsilon_\tau$ the naive
power-counting expectations are borne out: the random rod fixed
point (RR) (which becomes the usual Gaussian fixed point G at
$\epsilon = \epsilon_\tau = 0$) is stable against statistically
particle-hole symmetric perturbations ($g_0 = 0$, but $\Delta_g >
0$). However, it is strongly unstable to nonzero $g_0$, which
generates a crossover to a new, fully stable, particle-hole
asymmetric, or incommensurate (IC), critical fixed point (which
becomes a Bose Gaussian fixed point $\mathrm{G_B}$ at $\epsilon =
\epsilon_\tau = 0$).

However, for $\epsilon_\tau > \epsilon_\tau^c$ (with
$\epsilon_\tau^c = \frac{8}{29}$ to leading order in
$\epsilon_\tau$), nonlinear terms in the recursion relations
dominate the linear ones (the latter reflecting naive power
counting), and the RR fixed point becomes unstable to $\Delta_g$,
and a new statistically particle-hole symmetric, or commensurate
(C), fixed point bifurcates away from RR. For not too large
$\epsilon_\tau$ both C and RR remain unstable to $g_0$, and IC hence
remains the globally stable critical fixed point. However, for
$\epsilon_\tau > \epsilon_{\tau 1}$ (with $\epsilon_{\tau 1} =
\frac{2}{3}$ to leading order) the incommensurate fixed point
intersects the $g_0=0$ plane, and \emph{merges} with C. The latter
is now completely stable, and is proposed to correspond to the dirty
boson fixed point at $\epsilon_\tau = 1$. This provides a detailed
scenario by which statistical particle-hole symmetry is restored.
However, the fact that $\epsilon_\tau^c$, though small, is a finite
number means that the fixed point C is not perturbatively accessible
in the usual sense, and one should treat the detailed estimates in
the vicinity of this fixed point with caution. The value
$\epsilon_{\tau 1}$ is even larger, and therefore even more
uncertain, and extrapolation of these results to $\epsilon_\tau = 1$
should be treated as, at best, qualitative estimates. The general
scenario proposed, however, seems very natural and illuminating.

\section{Summary and conclusions}
\label{sec:conclude}

Over the past twenty years, advances in understanding the rich
physics of the phases and phase transitions in disordered boson
systems has proceeded fruitfully along a number of different fronts,
including descriptions of excitations within the different
phases,\cite{FWGF,WM08} scaling phenomenology for the critical
exponents\cite{FWGF,WM07} and universal
amplitudes,\cite{PHSa,PHSb,KW91,SWGY1992} exact solutions in one
dimension,\cite{Refaela,Refaelb,GS87,GS88,H1998} numerical
simulations in one and two
dimensions,\cite{SBZ1991,KTC1991,WSGY1994,ZKCG1995,AS2003,PS2004,Baranger06,SWGY1992}
weak hopping expansion work,\cite{FM1994,NFM1999} and approximate
renormalization group epsilon expansion approaches in higher
dimensions.\cite{WM08,MW96} I have tried here to give a flavor for
all of these, emphasizing qualitative features, such as phase
diagrams and the importance of particle-hole
symmetry,\cite{WM08,PHSa,PHSb,MW96} and making connections to
experiment\cite{Reppy1983,BEClatta,BEClattb,HLG1989,HP1990,GM1998}
where possible. Perhaps the most interesting future development
would be quantitative estimates of the BG--SF critical behavior in
three dimensions, which appears to be quantitatively accessible only
through numerical simulation. This would allow direct comparisons
with the Vycor data (at least for those exponents that appear to be
reasonably constrained by the data; of course, better experimental
helium data, perhaps using other disordered substrates, would be
hugely beneficial as well). However, this may have to await future
generations of computational capability, as the required system
sizes are currently well out of reach.

\end{document}